\documentclass[12pt]{article}
\usepackage{amssymb}
\def\inh{\vskip 0.075truein \noindent\hangindent=12 pt \hangafter=1}

\usepackage{amsthm}\theoremstyle{remark}

\newcommand{\bte}{\begin{quote}\begin{theorem}}
\newcommand{\ete}[1]{\label{#1}\end{theorem}\end{quote}}
\newcommand{\bcom}{\begin{quote}\end{quote}}
\newcommand{\bex}{\begin{quote}\begin{example}}
\newcommand{\eex}[1]{\label{#1}\end{example}\end{quote}}
\newcommand{\bcon}{\begin{quote}\begin{conclusion}}
\newcommand{\econ}[1]{\label{#1}\end{conclusion}\end{quote}}
\newcommand{\bdefi}{\begin{quote}\begin{definition}}
\newcommand{\edefi}[1]{\label{#1}\end{definition}\end{quote}}

\newcommand{\blem}{\begin{quote}\begin{lemma}}
\newcommand{\elem}[1]{\label{#1}\end{lemma}\end{quote}}

\newcommand{\bpr}{\begin{quote}\begin{problem}}
\newcommand{\epr}[1]{\label{#1}\end{problem}\end{quote}}

\newcommand{\f}{\frac}

\newcommand{\n}{\nonumber \\}
\newcommand{\inti}{\int_{-\infty}^\infty}
\newcommand{\beq}{\begin{eqnarray}}
\newcommand{\eeq}[1]{\label{#1}\end{eqnarray}}
\newcommand\eq[1]{(\ref{#1})}
\newcommand{\bfi}{\begin{figure}[24]}
\newcommand{\efi}[1]{\caption{\label{#1}}\end{figure}}
\newcommand\fig[1]{Fig.~\ref{#1}}
\newcommand{\res}{respectively}
\newcommand\gl{\left}
\newcommand\gr{\right}
\newcommand{\bfm}[1]{\mbox{\boldmath ${#1}$}}

\newcommand{\CE}{{\cal E}}

\newcommand{\CU}{{\cal U}}

\newcommand{\Ga}{\alpha}

\newcommand{\Gd}{\delta}
\newcommand{\Ge}{\varepsilon}

\newcommand{\Gg}{\gamma}

\newcommand{\Gk}{\varkappa}
\newcommand{\Gl}{\lambda}
\newcommand{\Gn}{\eta}
\newcommand{\Gm}{\mu}

\newcommand{\Gs}{\sigma}

\newcommand{\GD}{\Delta}
\newcommand{\GF}{\Phi}

\newcommand{\ind}[1]{{\it #1}\index{#1}}

\newcommand{\az}[1]{Sect.$\!$ \ref{#1}}
\newcommand\D{\,\mathrm{d}}
\newcommand\I{\mathrm{i}}
\newcommand\E{\mathrm{e}}
\newcommand{\bexe}{\begin{quote}\begin{exercise}\inh}
\newcommand{\eexe}[1]{\label{#1}\end{exercise}\end{quote}}

\usepackage{graphics,graphicx}
\usepackage{color}
\usepackage{epstopdf}
\begin{document}
{\large
\title{Brittle fracture in a periodic structure with internal potential energy}}

\author{Gennady S. Mishuris$^{a}$, and Leonid I. Slepyan$^{b,a}$}
\date{\small{ $^a${\em
Institute of Mathematics and Physics,
Aberystwyth University\\
Ceredigion SY23 3BZ
Wales UK} \\
$^b${\em School of Mechanical Engineering, Tel Aviv University\\
P.O. Box 39040, Ramat Aviv 69978 Tel Aviv, Israel}
}}

\maketitle

\vspace{3mm}\noindent
{\bf Abstract}
\noindent We consider a brittle fracture taking account of self-equilibrated distributed microstresses. To determine how the latter can affect the crack equilibrium and growth, a model of a structured linearly elastic body is introduced consisting of two equal symmetrically arranged layers (or half-planes) connected by an interface as a prospective crack path. The interface is comprised of a discrete set of elastic bonds. In the initial state with no external forces, the bonds are assumed to be stressed in such a way that tensile and compressive forces of the same value alternate. In the general considerations, the layers are assumed to be of an unspecified periodic structure, where such self-equilibrated stresses may also exist. A two-line chain and a lattice are examined as the specified structure. We consider the states of the body-with-a-crack under the microstresses and under a combined action of the remote forces and the microstresses. Analytical solutions to the considered problems are presented based on the introduction of a selective discrete transform. We show that the microstresses can manifest themselves in a number of phenomena such as the crack bridging, occurrence of `porosity' in front of the crack and crack growth irregularities. We demonstrate analytically and graphically that the microstresses can result in an increase as well as in a decrease of the crack resistance depending on the internal energy level. We also discuss different scenarios of the crack growth depending on the internal energy level. The results for the specified structure are presented with their dependence on  anisotropy, whose role in the considered problem appears substantial. The corresponding dynamic problems, the spontaneous crack propagation under the internal energy and the crack dynamics under the combined action of the internal energy and remote forces will be submitted separately.

\vspace{5mm}

\section{Introduction}
In this paper, we consider the {\em brittle fracture} with account of self-equilibrated microstresses associated with incompatible deformations alternating in the prospective crack path. In presence of such stresses, the growing crack releases some of the internal potential energy stored in a vicinity of the crack path. The questions are how much energy is released, what part of this energy goes to fracture itself and how the internal energy affects the crack resistance and the crack growth.

In this context, we recall that the theory of brittle fracture founded in 1920 by Griffith (1920, 1924) is based on the linear elasticity, in the framework of which the stress field is determined, and the energy criterion, which states that the energy release through the moving crack tip, must be equal to the (double) surface energy of the material.  It is important to note that the surface energy can be determined in an independent way. So the physical model by itself looks impeccable. Nevertheless, the  Griffith theory leaves some questions unanswered. One of them concerns the bulk-to-surface energy transition as the energy flux through the crack tip singular point (a line in the 3D case). In this theory, the mechanism of the transition is hidden, the microstructure role and the transition associated dynamic effects are not reflected.

Since then fracture mechanics  developed intensively in different directions, different materials and body shapes were considered, and different
modifications to the fracture criterion were introduced. In the context of the considered problem, the introduction of a natural unit of length is important. This was done in 1959 when, with the goal of bringing the transition mechanism to the macrolevel, Barenblatt (1959a,b, 1962) introduced a crack model with cohesive forces bridging the crack faces in a `small' autonomous region attached to the crack tip. In this model, the singular point is eliminated, and the fracture mechanism including the bulk-to-surface energy transition is transparent. Besides, a natural length unit introduced in this model can serve as the natural scale for the crack length. Note that the Griffith formulation follows from the Barenblatt model in the zero limit of the cohesive region length (in this connection, see Willis, 1967; Slepyan, 1981b, pp. 100-102; 2002, Sects. 5.10 and 7.4).

As a next step important for the present work, discrete models where introduced into considerations. In 1969 Novoghilov (1969a,b) formulated the concept of a brittle fracture, which took account of the discrete structure of the body, and suggested the necessary and sufficient criterion for the estimation of the strength of an elastic body weakened by a cut. The process of destruction is treated as a loss of stability of elastic equilibrium. These two factors, the discreteness and the loss of stability during the deformation of the breaking bond, are the basis of a number of phenomena that could not be detected in the framework of the continuum mechanics (see, e.g., Slepyan, 2010a). One of them is the energy radiation from the crack front in each act of the rupture (Thompson et al, 1971)).

The crack dynamics in a lattice model, for which both these factors are inherent, was first considered analytically in 1981 (Slepyan, 1981a). In this problem, the local-to-global energy release ratio was determined. This ratio plays the role of the crack-speed-dependent and material-structure-dependent corrective coefficient in the expression of the Griffith energy criterion. For the mass-spring square lattice it appeared that, in the quasi-static crack growth, only the proportion $\sqrt{2} - 1$  of the macrolevel energy release is spent on the fracture itself, while the remainder is radiated with the lattice waves. As the crack speed increases this ratio first increases not monotonically and then monotonically decreases (it tends to zero as the crack speed approaches the long wave speed). The structure of the radiation is described in more detail in Slepyan, 2010b.

In addition to the above-mentioned factors, the internal strain energy of self-equilibrated microstresses can play an important role. In addition to their effect on the crack resistance, some phenomena manifested in fracture, such as bridging, developing porosity in front of the crack and irregularities in crack growth can be caused by the microstresses.

Note that macrolevel residual stresses, self-equilibrated in a macro domain were considered repeatedly. In this connection, see the series of works by Banks-Sills et al (1997 $-$ 2006; also see references therein) and Bebamzadeh et al (2009, 2010)), where the role of curing residual stresses in the fracture of composites is examined. Such residual stresses manifest themselves as an additional load on the cracked body.

\begin{figure}[h!]
    \hspace{15mm}\includegraphics [scale=0.6]{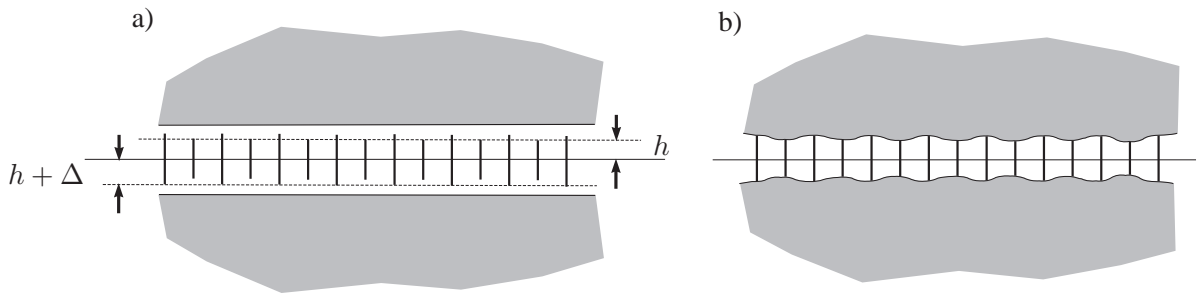}
    \put(-205,53){\small$h$}
    \put(-449,42){\small$h+\Delta$}
    \caption{The body with the structured interface. The compressed and stretched bonds alternate creating self-equilibrated stresses.}
\label{f1}
\end{figure}

We here consider a mechanism of fracture under the microstresses self-equilibrated in each cell of periodicity. A general formulation is used, where only the structure of the interface as the prospective crack path is specified.
The interface is assumed to be formed by a discrete set of uniformly distributed differently stressed bonds, where compressed bonds alternate with stretched ones. For example, this may happen if the initial lengths of the elastic bonds are different. The body is assumed to be symmetric about  the middle line crossing the bonds and periodic along this line, \fig{f1}. The  response of the structure to external action is reflected by means of a non-specified crack-related Green's function.

Note that a sketch of the corresponding dynamic problem was presented in the book by Slepyan (1981, pp. 272-275) in the framework of the elastic continuum with a crack subjected to a negative cohesive stresses.

We examine the crack equilibrium and discuss its spontaneous growth under the microstresses and then $-$ the crack equilibrium and growth under the combined action of the remote and micro stresses.
During the crack growth the internal potential energy is released in part, and one of the tasks of the present study is to determine what proportion of the released energy is going to fracture and what part is radiated. Another question is how the microstresses affect the crack equilibrium and the crack growth mode. Note that in some respects the problem is related to the phase transition in a lattice (Slepyan and Ayzenberg-Stepanenko, 2002), to the bridge crack (Mishuris et al, 2008) and to the fracture of a structural interface (Mishuris et al, 2012).

In the considered problem, there is no regular steady-state limit but there exists two-bond clustering. Under the condition of the non-specified structure, this requires the {\em selective} discrete transform, that is the Fourier discrete transform on the related points, one in each cluster. The expression of the latter transform in terms of the regular discrete transform is presented. Analytical techniques allow us to find the final solution based on the general formulation, whereas numerical illustrations can be obtained from this as soon as the Green's function is specified. For this goal we use a two-line chain and the 2D square lattice.

We determine the energy relations and discuss possible scenarios of the crack growth, which are defined by the bond length ratio and the ratio of the internal energy to the energy coming from the remote load. The role of the dynamic factor (Slepyan, 2000) in the fracture under the microstresses is discussed. It is found, in particular, that under the microstresses the crack initiation energy barrier can increase, whereas the crack growth can be accompanied by irregularities and clustering. In some respect, these phenomena are similar to those found earlier for mode II crack dynamics in a triangular lattice (Slepyan and Ayzenberg-Stepanenko, 2002) and in a lattice under the harmonic excitation (Mishuris et al, 2009, and Slepyan et al, 2010).

The paper is organised as follows. First a simple model is considered consisting of two parallel elastic strings connected by a discrete set of periodically placed bonds alternating by their initial lengths. The structure is under the microstresses, no external forces are applied. It is assumed that under this condition only initially stretched bonds may break. In spite of the simplicity, this model demonstrates all main effects due to microstresses presence. We determine the initial prestress \eq{aie2b}, the tensile forces in the crack front bond in the case of a semi-infinite bridged crack \eq{aie7force} and for only one \eq{ef3Q} or two bonds broken \eq{u10p}. Also the corresponding energy relations, as the ratios of the energy of the bond to the initially stored or released energy, are presented  \eq{ieoab2}, \eq{efer}, \eq{R0to14} and \eq{1to2t3}. Here and below the results are presented in dependence on the orthotropy parameter, $\Ga$, as the ratio of the bond stiffness to that of a string between-the-bonds segment.

Next a more general structure is considered where only the crack path structure, the same as for the chain, is specified, whereas the bulk of the body is reflected by the crack-related non-specified Green's function. The same values are determined in a general form as for the chain with the semi-infinite bridge crack. The general results are expressed through the Wiener-Hopf equation kernel, which, in its turn, is expressed through the Green's function by means of the {\em selective} discrete transform introduced, \eq{ss2}, \eq{geo1} and Appendix. The numerical results for a specific structure can be obtained as soon as the Green's function is given. This is done for the chain and the lattice.

Then, we consider the problem in the general formulation, where both types of actions are present: the internal prestresses and external forces. Two different external loads are considered, one uniformly distributed at infinity and the other as a remote force defined by the corresponding stress intensity factor. A semi-infinite open crack  without the crack-face bridging corresponds to the latter. In the latter case, the tensile forces are determined for the crack front bond and the next one \eq{suca66}. This allows us to reveal different scenarios of the crack growth corresponding to different levels of the internal energy. It is found, in particular, that in an initial plot of the internal energy level, the resistance to the crack growth (the macrolevel energy release rate) increases, then it decreases, \eq{suca131}, \eq{suca131a}, \fig{f11}, \fig{f12} and \az{CS}. Respectively the fracture regime is changed.

Finally, the discussions and conclusions are presented.

\subsection{The energies and displacements in the considered problem}
\subsubsection{The energies}
The state of the considered structure is characterised by three values of the energy. The first is the {\em initial internal energy}, $\CE$, which arises due to the difference in the interface bond lengths and is stored in the cell of periodicity, that is, in two spans of the structure.  Next is the {\em initial energy of the bond}, $E$, which is the same for all the bonds. These two values correspond to the initial state of the intact structure. Lastly, the {\em actual energy of the bond}, $E_m$, where $m$ is the bond number, is considered. These energies refer to the microlevel. The critical values of the initial internal energy and the bond energy are denoted by $\CE_c$ and $E_c$, \res, where the latter is also the same for all the bonds (in the initial state, the critical value of $\CE=\CE_c$ corresponds to the critical value of the bond energy, $E_c$). The level of the internal energy is characterised by the {\em ratio of the stored internal energy to its critical value}, $\Gg=\CE/\CE_c=E/E_c$.

In a steady-state crack growth, the  energy $\CE$ is released, so the microlevel energy release rate $G_{mic} = \CE/(2a)$, where $a$ is the between-the-bond distance. A part of this energy disappears with the breaking bond, while the other part is radiated in the form of acoustic oscillations. Due to linearity of the problem, $\CE$ is proportional to $\GD^2$, where $\GD$ is a half of the difference in the lengths of the unstrained bonds. The analysis is given below mainly in terms of $\GD$. If the internal energy is given instead of the bond length initial difference, $2\GD$, the results can be read based on the relation between the initial internal energy and $\GD$.

In the case where a remote external load is present, its action is reflected, as usual, by the macrolevel energy release rate, $G_{mac}$. In this case, we examine how the internal energy affects the critical value of $G_{mac}$.

\subsubsection{The position coordinates and displacements}
In the $x,y$ rectangular coordinate system, where $y=0$ is the axis of symmetry, $x=0$ for the bond $m=0$, $y=\pm h$ for the even number unstressed bond ends and $y=\pm (h+\GD)$ for the odd number bonds. The displacements in $y$-direction, $\pm u_m$, correspond to the upper and lower ends of the $m$-bond, \res. In these terms, in the initial state of the structure, the displacements are nonzero as well as the tensile forces. We denote the initial and additional displacements by $\CU_m$ and $U_m$, \res, so that the actual (total) displacements $u_m=\pm \CU_m +U_m$. Note that in the initial state the bonds are stressed equally differing only in the sign for the even and odd bonds, and
 \beq u_m = \CU_0 +U_m~~(m=0, \pm 2, ...)\,,~~~u_m= -\CU_0+U_m~~(m=\pm 1, \pm 3, ...)\,.\eeq{UandU}
The corresponding tensile forces are
 \beq Q_m = 2\Gk u_m\,,\eeq{ie4}
where $\Gk$ is the bond stiffness.

Concerning the displacements, we consider below only the upper end of the bond interacting with the respective part of the body. However, both parts of the body including the interface are taken into account when calculating the energy.

We start with a simple example of a two-line chain, where detailed solutions are achieved without using sophisticated mathematics.
Along with this, the transparent results obtained allow the main effects introduced by microstresses to be seen.
Then the problem is considered in a more general formulation; however, we also return to the chain to demonstrate how the results for the chain can be obtained as a particular case of the general solution. An orthotropic lattice is also used for this purpose.

\section{Two-line chain}\label{aie}
\subsection{The initial state}
\begin{figure}[h!]
    \hspace{6mm}
    \begin{picture}(0,0)(-10,45)
    \includegraphics [scale=0.50]{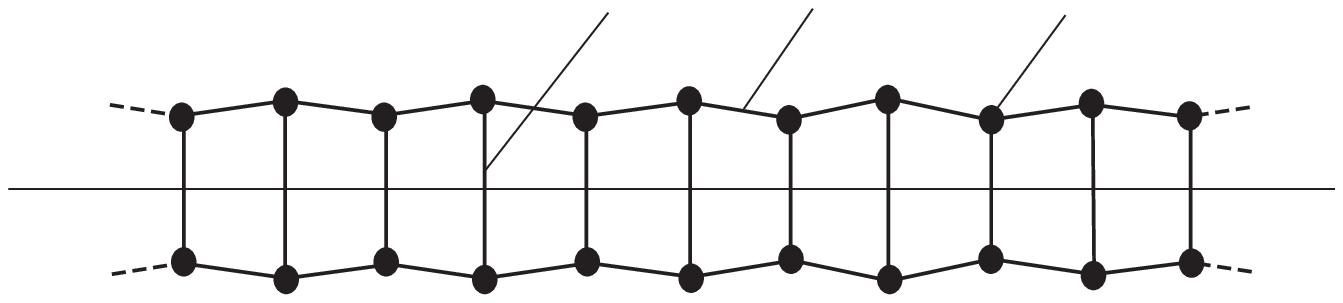}\hspace{10mm}
    \put(17,-13){\small$m=-6$\hspace{0.5mm}$-5$\hspace{0.5mm}$-4$\hspace{0.2mm}$-3$%
    \hspace{0.1mm}$-2$\hspace{0.5mm}$-1$\hspace{3mm}$0$\hspace{3.2mm}$1$\hspace{3.3mm}$2$\hspace{3.3mm}$3$\hspace{3.2mm}$4$}
    \put(20,46){\small b)}
    \end{picture}


    \begin{picture}(0,60)(-250,-15)
  \includegraphics [scale=0.50]{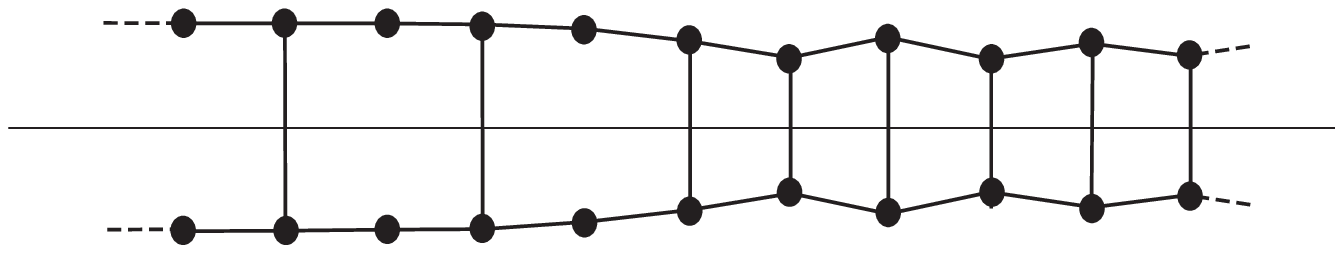}
    \put(-421,-13){\small$m=-4$\hspace{0.5mm}$-3$\hspace{0.5mm}$-2$\hspace{0.5mm}$-1$%
    \hspace{2.7mm}$0$\hspace{3mm}$1$\hspace{3mm}$2$\hspace{3.2mm}$3$\hspace{3.3mm}$4$\hspace{3.3mm}$5$\hspace{3.2mm}$6$}

\put(-415,46){\small a)}
\put(-321,46){\small$\varkappa$}
\put(-290,46){\small$\mu$}
\put(-256,46){\small$M$}

\end{picture}
    \caption{The chain structure: the intact chain (a) and the chain with a semi-infinite bridged crack (b). }
\label{f2}
\end{figure}

Consider two parallel strings connected by periodically placed bonds numbered by $m=0, \pm 1, ...$ , \fig{f2}a. Let the initial length of the even bonds be $2h$, whereas the odd bonds, $m=\pm 1, \pm 3, ...,$ be of a slightly different length, $2h+2\GD$. In the framework of the Hooke's law, the bond's stiffness, $\Gk$, is assumed to be the same for both the even and the odd bonds. The stiffness of the string section between the neighboring bonds is denoted by $\Gm$. Initially, when all the bonds are intact, the even and the odd bonds are deformed uniformly, that is, the positions of the bond's upper ends are $y_{2m}=y_0=h+\CU_0$  and $y_{2m+1}=y_1=h+\GD+\CU_1$, where $\CU_{0,1}$ are the displacements relative to the unstressed bond positions. It follows that the equilibrium equations are
 \beq (\Gm +\Gk)y_0 -\Gm y_1=\Gk h\,,~~~(\Gm +\Gk)y_1-\Gm y_0 = \Gk(h+\GD)\,.\eeq{aie11}
We find the displacements
 \beq \CU_0 = - \,\CU_1= \f{\GD}{2+\Ga}\,,~~~ \Ga=\f{\Gk}{\Gm}\,,\eeq{aie2a}
and the tensile forces in the bonds
 \beq Q_0=-Q_1 = 2 \Gk \CU_0\,.\eeq{aie2b}
Thus, the vertical coordinates of the stressed bonds are
 \beq y_0= h+\f{\GD}{2+\Ga}\,,~~~y_1=h+\GD-\f{\GD}{2+\Ga}= h+\f{1+\Ga}{2+\Ga}\GD\,.\eeq{inpos}
Recall that the stiffness of both the even and the odd bonds is assumed to be identical. Therefore, in terms of the displacements, the intact structure can be considered as uniform. If the odd bonds are longer than the even ones then
the latter are under tension, and some of them can be broken. The question is whether such
damage can spread. This question is considered below.

\vspace{5mm}

\subsection{A semi-infinite damaged region}
Here we consider the case where all the even bonds at the left, $m=-2, -4, ...$, are broken, see \fig{f2}b. The task is to find the tensile force, $Q_0$, at the front of this bridge crack. In the left region, the equilibrium equation is
 \beq
 u_{m-3} + u_{m+1} - 2(1 +2\Ga)u_{m-1} = 0\,,~~~(m=-2, -4, ...)\,,\eeq{aie3}
and the displacements can be represented as
 \beq u_{m+1} = U_{-1}\Gl_-^{-(m+2)}~~~(m=-2,-4,...)\,,\eeq{aie4}
 with
\beq \Gl_-^2 = 1+2\Ga -2\sqrt{\Ga +\Ga^2}\,.\eeq{aie5}
Note that the expression for $\Gl_-$ follows from \eq{aie3} and the boundedness condition at minus infinity (in fact, $u_{-\infty}=0$).
The equilibrium equation with respect to the displacements in the intact bond region, $U_{m}$, $m=0,1,2,...$, {\em additional} to those defined in the intact structure \eq{inpos} is
 \beq U_{m+1}+U_{m-1} - 2(1+\Ga)U_m =0~~~(m=1,2, ...)\,,\eeq{aie6}
and the total displacements in this region can be represented as (see \eq{aie2a})
 \beq u_{2m}= \f{\GD}{2+\Ga} +U_0\Gl_+^{2m}\,,~~~ u_{2m+1} = - \f{\GD}{2+\Ga} + U_0\Gl_+^{2m+1}~~~(m=0, 1,2,...)\,,\n
 \Gl_+=1+\Ga-\sqrt{2\Ga+\Ga^2}\,.\eeq{aie7}

The remaining values, $U_0$ and $U_{-1}$, are defined by the equilibrium equations with respect to points $m=0$ and $m=-1$
 \beq y_{-1}+y_1-2y_0 - 2\Ga(y_0-h) =0\,,\n
\f{1}{2}(y_{-3}-y_{-1})+y_0-y_{-1} -2\Ga(y_{-1}-h-\GD) =0\,.\eeq{aie5a}
From Eqs. \eq{aie4} and \eq{aie7} it follows that
 \beq y_{-3}= h+\GD+U_{-1}\Gl_-^2\,,~~~y_{-1}=h+\GD+U_{-1}\,,\n y_0= \f{\GD}{2+\Ga}+h+U_0\,,~~~ y_1=h+\f{(1+\Ga)\GD}{2+\Ga}+U_0\Gl_+\,,\eeq{aie6a}
and the equations \eq{aie5a} can be rewritten in the form as
 \beq (2+2\Ga-\Gl_+)U_0 - U_{-1}  = \f{\GD}{2+\Ga}\,,\n
2U_0 - (3+4\Ga-\Gl_-^2)U_{-1} = \f{2(1+\Ga)\GD}{2+\Ga}\eeq{aie7a}
with the tensile force at the bridge crack front, $m=0$
 \beq Q_0 =2\Gk\gl(\f{\GD}{2+\Ga}+U_0\gr)\,.\eeq{aie7force}

\subsection{One-two even bonds are broken}
\begin{figure}[h!]
    \hspace{6mm}
    \begin{picture}(0,0)(-10,53)
    \includegraphics [scale=0.50]{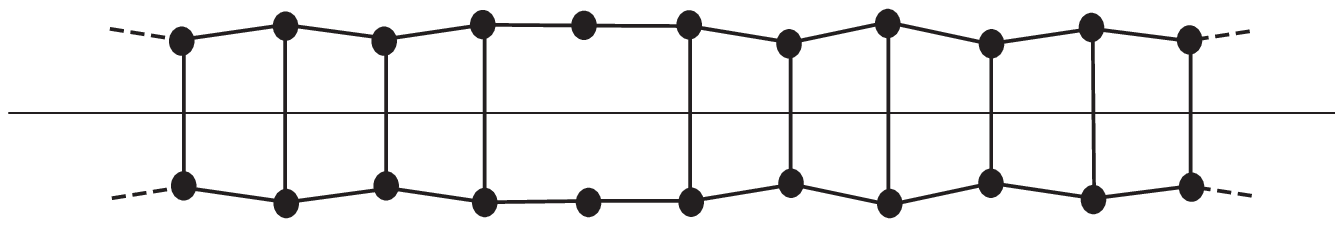}\hspace{10mm}
    \put(20,46){\small b)}
    \put(17,-13){\small$m=-4$\hspace{0.5mm}$-3$\hspace{0.5mm}$-2$\hspace{0.5mm}$-1$%
    \hspace{2.7mm}$0$\hspace{3mm}$1$\hspace{3mm}$2$\hspace{3.2mm}$3$\hspace{3.3mm}$4$\hspace{3.3mm}$5$\hspace{3.2mm}$6$}
    \end{picture}

    \begin{picture}(0,70)(-250,-20)
  \includegraphics [scale=0.50]{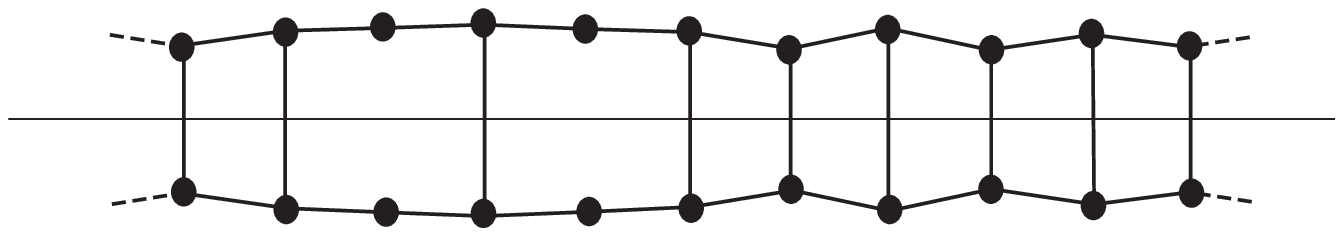}
     \put(-421,-13){\small$m=-4$\hspace{0.5mm}$-3$\hspace{0.5mm}$-2$\hspace{0.5mm}$-1$%
    \hspace{2.7mm}$0$\hspace{3mm}$1$\hspace{3mm}$2$\hspace{3.2mm}$3$\hspace{3.3mm}$4$\hspace{3.3mm}$5$\hspace{3.2mm}$6$}
    \put(-415,46){\small a)}
    \end{picture}

    \caption{The chain with one and two initially stretched bonds broken}
\label{f4}
\end{figure}

Let only the `central' bond, $m=0$, be broken, \fig{f4}a.  Since, in this case, $y_{-1}=y_1=y_0$ the equilibrium equation with respect to the point $m=1$ is
 \beq y_2-y_1-2\Ga(y_1-h-\GD)=0\eeq{ef1}
with
 \beq y_1=h+\f{(1+\Ga)\GD}{2+\Ga}+U_1^{(1)}\,,~~~y_2=h+\f{\GD}{2+\Ga}+U_1^{(1)}\Gl_+\,,\eeq{ef2}
where $U_1^{(1)}$ is the displacement {\em additional} to that defined in the intact structure \eq{inpos}. It follows that
 \beq U_1^{(1)}= \f{\Ga\GD}{(2+\Ga)(1+2\Ga-\Gl_+)}\,,\n u_1= u_1^{(1)}=-\f{\GD}{2+\Ga}\gl(1-\f{\Ga}{1+2\Ga-\Gl_+}\gr)\,,\n
 u_2=u_2^{(1)}=\f{\GD}{2+\Ga}\gl(1+\f{\Ga\Gl_+}{1+2\Ga-\Gl_+}\gr)\,,\eeq{ef3}
and due to the symmetry
 \beq u_{-1}=u_0=u_1\,,~~~u_{-2}=u_2\,.\eeq{ef33}
Here and below, in this Section, the number of broken bonds is shown in brackets in the superscript. Thus, the tensile force at the bridge crack fronts, $m=\pm 2$ is
\beq Q_2^{(1)}=\f{2\Gk\GD}{2+\Ga}\gl(1+\f{\Ga\Gl_+}{1+2\Ga-\Gl_+}\gr)\,,\eeq{ef3Q}

In a similar way, for the case where two bonds, $m=0$ and $m=-2$, are broken, \fig{f4}b, we have two equations for the {\em additional} displacements
 \beq (1+2\Ga)U_{-1}^{(2)} - U_1^{(2)} = \f{2\Ga\GD}{2+\Ga}\,,\n
 U_{-1}^{(2)}- (3+4\Ga-2\Gl_+)U_1^{(2)} =- \f{2\Ga\GD}{2+\Ga}\,.\eeq{ef4}
Thus, the total displacements at points $m=-1,1,2$ are, respectively
 \beq u_{j}^{(2)}=-\f{\GD}{2+\Ga}+U_j^{(2)}\,,~~( j=-1,1)\,,~~~ u_2^{(2)}=\f{\GD}{2+\Ga}+U_1^{(2)}\Gl_+\eeq{ef5}
with
 \beq U_1^{(2)}=\f{4\Ga(1+\Ga)\GD}{(2+\Ga)[(1+2\Ga)(3+4\Ga-2\Gl_+)-1]}\,,~~~Q_2^{(2)}=2\Gk u_2\,,\eeq{u10p}
 where $U_{-1}^{(2)}$ to be found from \eq{ef4}.

\begin{figure}[h!]
  \begin{center}
    \includegraphics [scale=0.50]{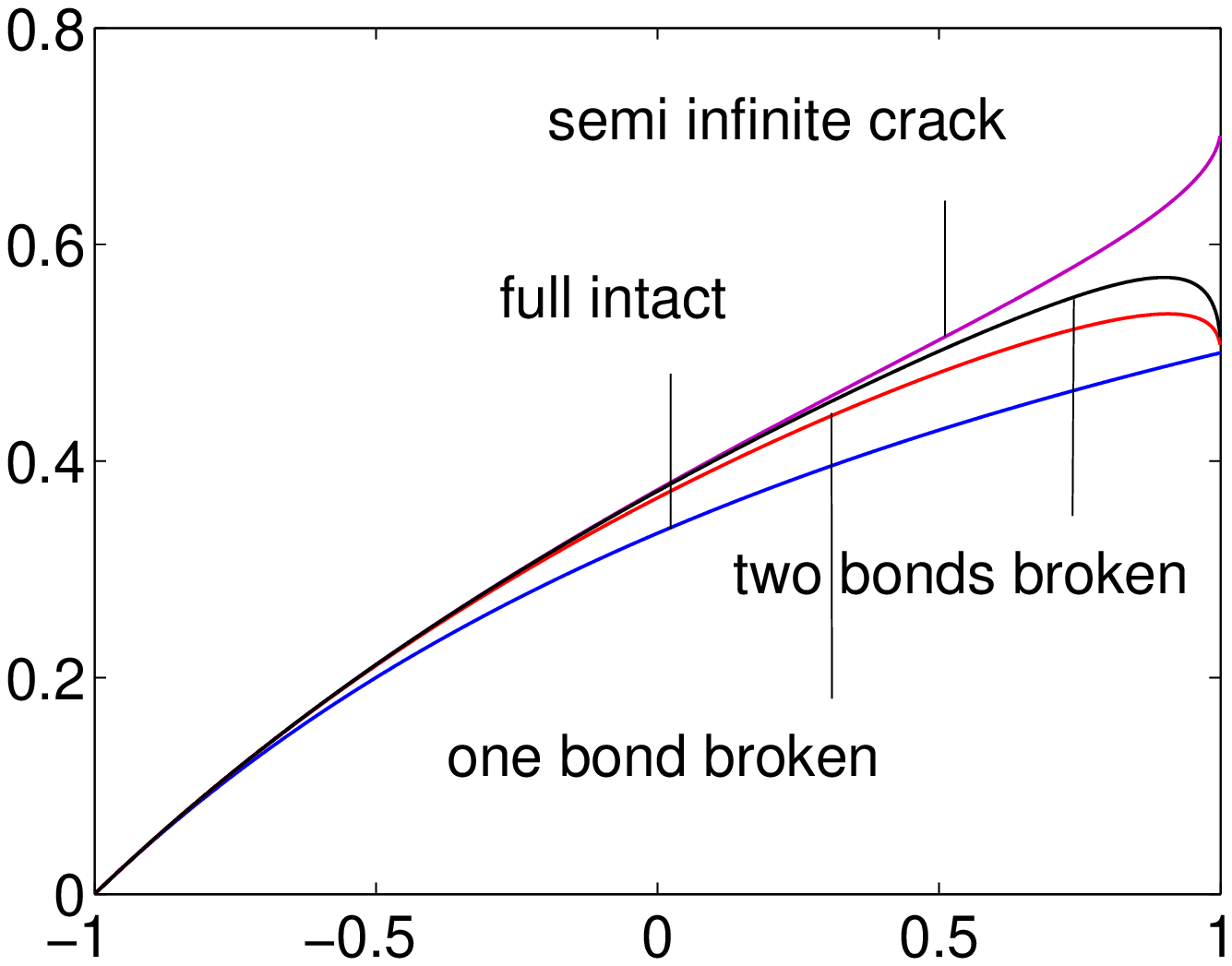}
    \put(-250,90){$\displaystyle\frac{Q_{max}}{Q^0}$}
       \put(-112,-10){$\hat\alpha$}
  \end{center}
    \caption{Normalised maximal tensile forces, $Q_{max}/Q^0$, for the four considered states shown in \fig{f2} and \fig{f4} ($Q^0=2\Gk\GD, \hat\alpha=(1-\Ga)/(1+\Ga)$).}
\label{f5}
\end{figure}

The normalised maximal tensile forces corresponding to the initial and the three other states considered above are plotted in \fig{f5} as a function of $\hat{\Ga}=(1-\Ga)/(1+\Ga)$. The calculations are based on the relations in \eq{aie2b}, \eq{aie7force}, \eq{ef3Q} and \eq{u10p}. It can be seen that for not too small $\Ga$ the maximal value of the tensile force approaches that for the semi-infinite crack very fast.  Practically, the curves for one-two broken bonds almost coincide with the limiting one.
Below, in \az{aie1}, we return to this example considering it in a very different way by following a more general formulation.

\subsection{Energy relations and the crack initiation criterion}
The strain energy per two spans stored in the initial state of the chain is
 \beq \CE = 2\Gm(y_1-y_0)^2 + 4\Gk \CU_0^2 = \f{2\Gk\GD^2}{2+\Ga}\,,\eeq{ef6}
whereas the energy of the bond (the same for the even and the odd ones) is
 \beq E= 2\Gk \CU_0^2 =  \f{2\Gk\GD^2}{(2+\Ga)^2}\,.\eeq{ieoab1}
Thus, the initial energy ratio
 \beq R_0=\f{E}{\CE} = \f{1}{2+\Ga} \eeq{ieoab2}
evidences that the critical stored energy is sufficient for both bonds to break even at $\Ga=0$, and it increases with the parameter $\Ga$. In fact, however, the former is also spent for dynamic effects accompanying the breakage, and if $\CE$ is not too close to the critical value (where the bond's energy is critical) only even bonds are expected to break.

In the case of the semi-infinite crack, the strain energy in the crack front bond is
 \beq E_0=E_0^{(\infty)}=2\Gk \gl(\f{\GD}{2+\Ga}+U_0\gr)^2\,,\eeq{ef60}
where $U_0$ is defined by Eqs. \eq{aie7a}. The energy ratio is
 \beq R_\infty = \f{E_0^{(\infty)}}{\CE}=\f{1}{2+\Ga}\gl(1+\f{(2+\Ga)U_0}{\GD}\gr)^2\,.\eeq{efer}

For the transition from the initial intact state to that with one bond broken,  the energy release, $G_{0\to 1}$, can be calculated as a sum of the broken bond energy and the (double) static work of the force $Q_{0}$ on the additional displacement at $m=0$. With reference to \fig{f4}a we have
 \beq 2aG_{0\to 1} = E_0^{(0)} + Q_0\gl(y_1^{(1)}-y_0^{(0)}\gr)\,,~~~E_0^{(0)}=E=2\Gk \f{\GD^2}{(2+\Ga)^2}\,,\eeq{R0to1}
with
 \beq Q_0=\f{2\Gk\GD}{2+\Ga}\,,~~~y_0^{(0)} =h +\f{\GD}{2+\Ga}\,, ~~~y_1^{(1)}= h+\GD-\f{\GD}{2+\Ga}\gl(1 -\f{\Ga}{\Ga +\sqrt{2\Ga+\Ga^2}}\gr)\,.
 \eeq{R0to12}
It follows that the released energy
 \beq  2aG_{0\to 1} =\f{2\Gk\GD^2}{2+\Ga}\gl(1-\f{\sqrt{2\Ga+\Ga^2}}{(2+\Ga)(\Ga+\sqrt{2\Ga+\Ga^2})}\gr)\eeq{R0to13}
is less than the stored energy $\CE$ \eq{ef6}. In this case, in contrast to the case of the semi-infinite crack, the corresponding energy ratios of the bond energy, $E$, to the released energy and to the stored one are different
 \beq P_0 =\f{E_0^{(0)}}{2aG_{0\to 1}}=\f{\Ga +\sqrt{2\Ga+\Ga^2}}{[1+\Ga+\sqrt{2\Ga+\Ga^2}]\sqrt{2\Ga+\Ga^2}}\,,~~~R_0=\f{E_0^{(0)}}{\CE}=\f{1}{2+\Ga}<P_0\,,\eeq{R0to14}
where the superscript indicates the number of initially broken bonds.

In the similar way, calculations for the transition from the state with the only bond broken to that with two broken bonds lead to the following expression for the released energy
 \beq 2aG_{1\to 2} = E_{-2}^{(1)} + Q_{-2}^{(1)}(u_{-2}^{(2)}-u_{-2}^{(1)})=2\Gk u_{-2}^{(1)}u_{-2}^{(2)}\,,\eeq{1to2t1}
The energy ratios of the bond energy, $E_{-2}^{(1)}$, to the released energy and to the stored one, are
 \beq P_1 =  \f{E_{-2}^{(1)}}{2aG_{1\to 2}}\,,~~~R_1 = \f{E_{-2}^{(1)}}{\CE}\,.\eeq{1to2t3}
Recall that $E_{-2}^{(1)}=2\Gk (u_{-2}^{(1)})^2$.

\begin{figure}[h!]
  \begin{center}
    \hspace{15mm}\includegraphics [scale=0.50]{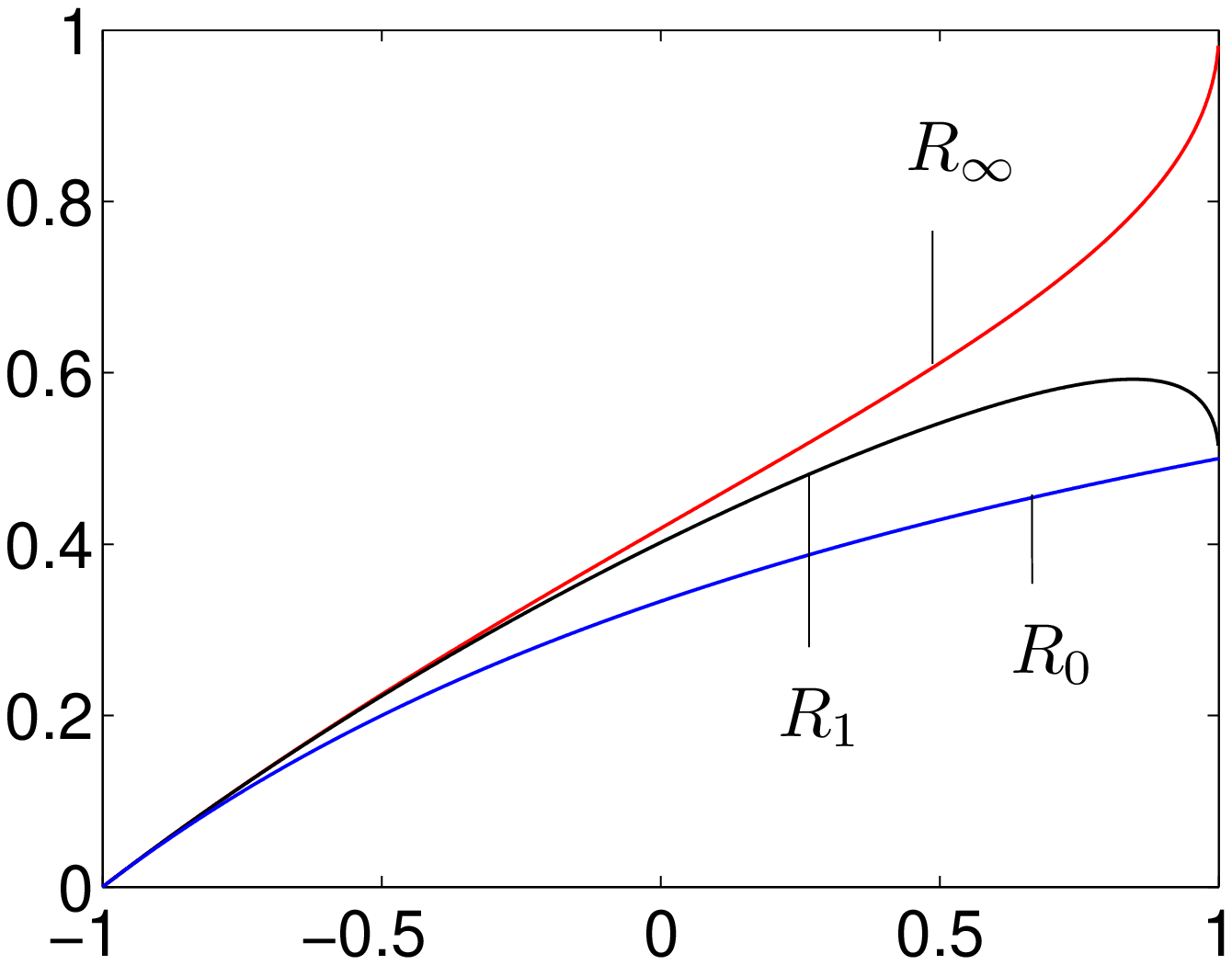}
        \put(-110,-10){\small$\hat\alpha$}
  \end{center}
    \caption{The energy ratios, $R_\infty$ \eq{efer}, $R_0$  \eq{R0to14} and $R_1$ \eq{1to2t3}, of the maximal bond energy to the stored energy as functions of the dimensionless parameter $\hat \alpha$ for the semi-infinite bridge crack and one and two broken bonds, \res.}
\label{f6}
\end{figure}

\begin{figure}[h!]

\begin{center}
   \includegraphics [scale=0.44]{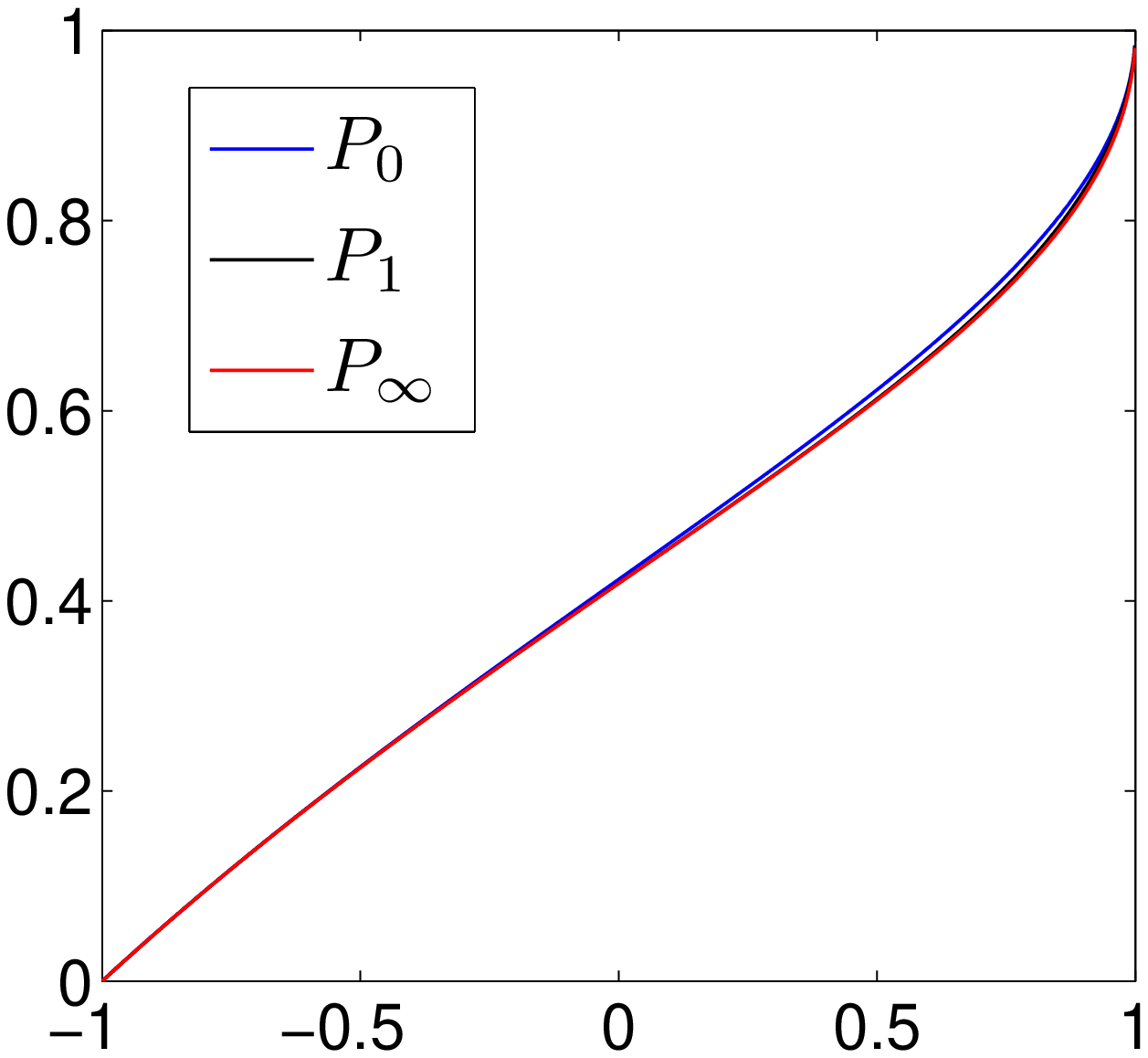}\hspace{10mm} \includegraphics [scale=0.50]{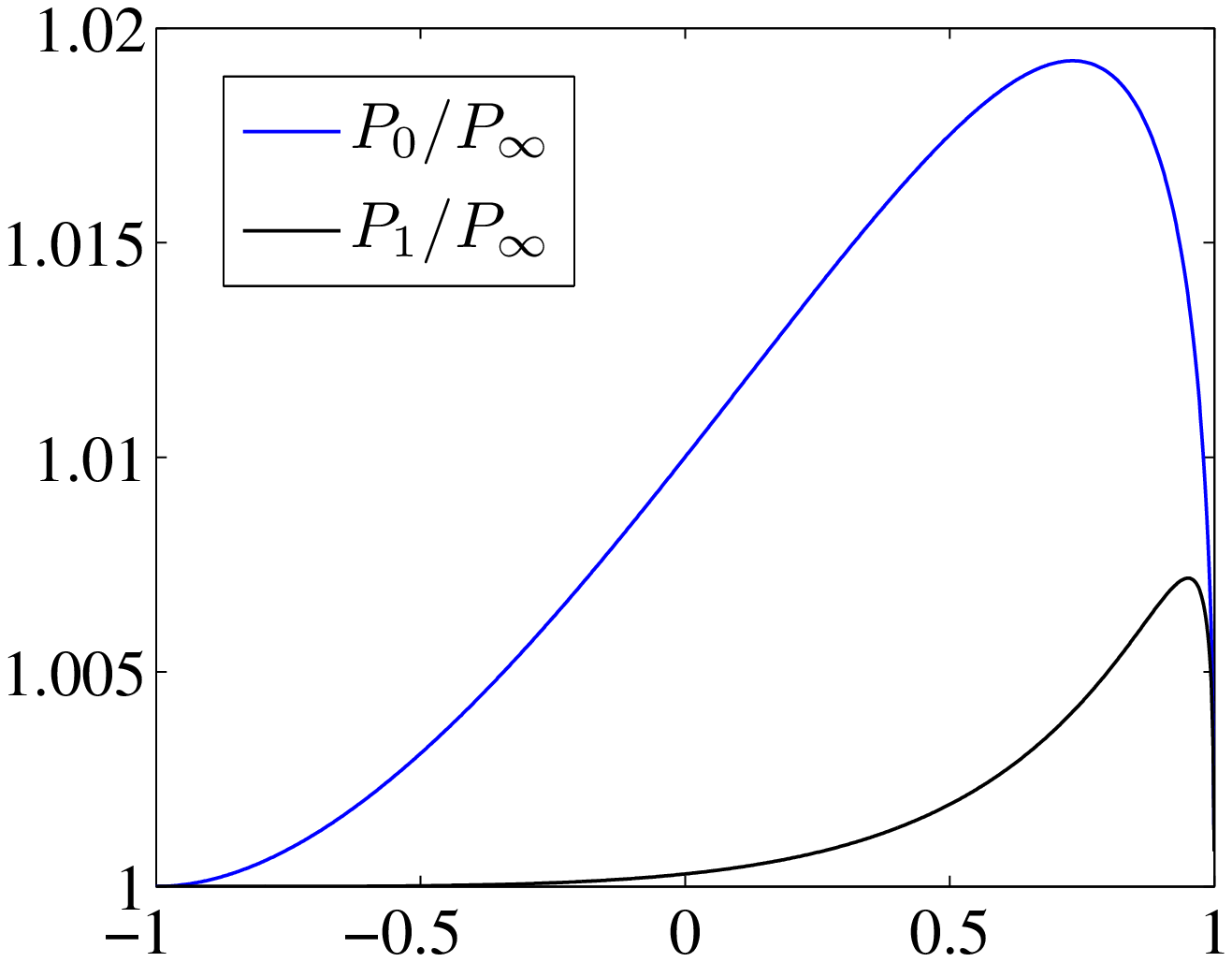}
        \put(-345,-7){\small$\hat\alpha$}
           \put(-110,-7){\small$\hat\alpha$}
\end{center}

    \caption{The energy ratios, $P_\infty=R_\infty<P_1<P_0$ of the maximal bond energy to the released energy as functions of the dimensionless parameter $\hat \alpha$ for the semi-infinite bridge crack \eq{efer} and one \eq{R0to14} and two \eq{1to2t3} broken bonds, \res, are presented in the left-side plot, which evidences that these three ratios are almost the same. Their ratios, $P_0/P_\infty$ and $P_1/P_\infty$, are shown in the right-side plot.}
\label{f7}
\end{figure}

The energy ratios for the semi-infinite bridge crack and one and two broken bonds, $R_\infty, R_0$ and $R_1$, \res, are presented in \fig{f6} as functions of $\hat{\Ga}$. The plots of the ratios $P_\infty = R_\infty, P_0$ and $P_1$ are shown in \fig{f7}. It is remarkable that the ratio of the fracture energy to the total released energy is practically the same for any finite and semi-infinite cracks.

In the energy terms, the crack initiation criterion is
 \beq \CE =E_c/ R_j\,,~~~j=0,1,\infty\,.\eeq{efec}
Recall that $E_{c}$ is the critical energy of the initially stretched bond and $\CE$ is the internal energy per two spans stored in the initial state of the chain. The factor $1/R_j$ defines how much internal energy is required for the next step of the crack advance. A part of this energy goes on fracture itself, that is for the bond breakage. The
remainder is radiated. The radiated acoustic energy, as the difference between the energy release and the critical energy of the bond, is equal to
 \beq  E_c \f{1-R_j}{R_j}\,,~~~j=0,1,\infty\,.\eeq{efecrad}
The radiated oscillations, however, act on the next initially stretched bond and this can lead to the dynamic crack growth even in the case where the stored energy is below the critical level defined in \eq{efec} (in this connection, see Slepyan, 2000).  Thus, the crack propagation criterion appears below the crack initiation threshold as is usually observed.

\subsubsection{The limiting initial energy and two other limiting relations}\label{lietlr}
Let $\Gg$ be the ratio of the stored internal energy, $\CE$, to its critical value, $\CE_c$, corresponding to the critical initial state of the bond: $E=E_c$ as $\CE=\CE_c$. In the initial state, the former equality is achieved for $\Gg=1$, whereas in the presence of a crack, it is satisfied at the crack front bond for $\Gg=\Gg_c<1$. We now calculate $\Gg_c$ and its limiting values (for $\Ga\to 0$ and $\Ga\to\infty$) corresponding to the critical energy of the first even bond, $m=0$, in front of the semi-infinite crack bridged by the odd bonds, $m=-1, -3, ...$ . Referring to \eq{ef6} we have
 \beq \CE = \f{2\Gk\GD^2}{2+\Ga} = \Gg\CE_c\,,\eeq{tlrs1}
where $\CE_c$ is the critical internal energy. At the same time, the energy of the bond is
 \beq E = 2\Gk \CU_0^2=\f{\CE}{2+\Ga}\,,~~E_c = \f{\CE_c}{2+\Ga}\,,~~\CU_0=\f{\GD}{2+\Ga}~~~(\mbox{in the initial state})\,,\n
 E_0 = 2\Gk \gl(\CU_0+U_0\gr)^2~~~(\mbox{in the actual state with a semi-infinite bridged crack})\,.\eeq{tlrs2}
Now for $E_0=E_c$ we get
 \beq 2\Gk\gl(\CU_0+U_0\gr)^2 = E_c=\f{\CE_c}{2+\Ga}=\f{\CE}{(2+\Ga)\Gg}=2\Gk\f{\CU_0^2}{\Gg}\,,\eeq{N2}
and we obtain from this that
 \beq \Gg = \Gg_c = \gl(1+\f{U_0}{\CU_0}\gr)^2\,.\eeq{tlrs4}

Next, as follows from \eq{ef4}
 \beq \lim_{\Ga\to 0}U_0= \f{\GD}{2(1+\sqrt{2})}\,,~~~\lim_{\Ga\to \infty}\Ga U_0=0\,;\eeq{tlrs5}
hence
 \beq  \lim_{\Ga\to 0}\Gg_c= 1/2\,,~~~\lim_{\Ga\to \infty}\Gg_c=1\,.\eeq{tlrs6}

\begin{figure}[h!]
  \begin{center}
    \hspace{15mm}\includegraphics [scale=0.50]{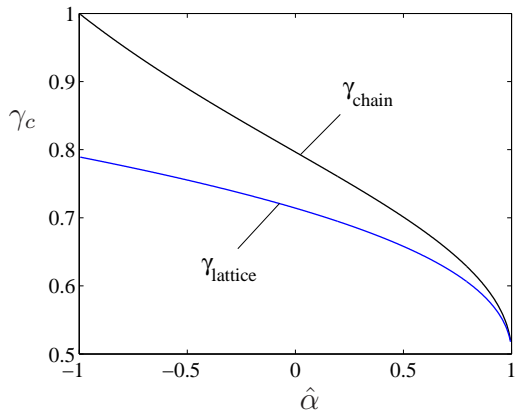}
        \put(-210,105){$\displaystyle\gamma_c$}
       \put(-100,-3){$\hat\alpha$}
  \end{center}

  \vspace{-5mm}
    \caption{The internal energy critical level, $\Gg_c$, for the chain and the lattice with a semi-infinite bridged crack, as functions of the dimensionless parameter $\hat \alpha$.}
\label{Gg}
\end{figure}

Plots of $\Gg$ for the chain and the lattice considered in \az{scl} are presented in \fig{Gg}.

\section{A general structure}
\subsection{The initial state}
We consider a 2D non-specified structure consisting of two equal periodic half-planes or layers connected by a set of elastic bonds, \fig{f1}b. The  bonds are numbered by $m = 0, \pm 1, ...$ .  As above the even and odd bonds differ only by their initial length, namely, the even bonds, $m=0, \pm 2, ...,$ are of the length $2h$, whereas the odd bonds, $m=\pm 1, \pm 3, ...,$ are of a slightly different length, $2h+2\GD$. In the framework of the  Hooke's law, the bond's stiffness, $\Gk$, is assumed to be the same for both the even and the odd bonds.

\begin{figure}[h!]

  \vspace{5mm}
    \hspace{15mm}\includegraphics [scale=0.6]{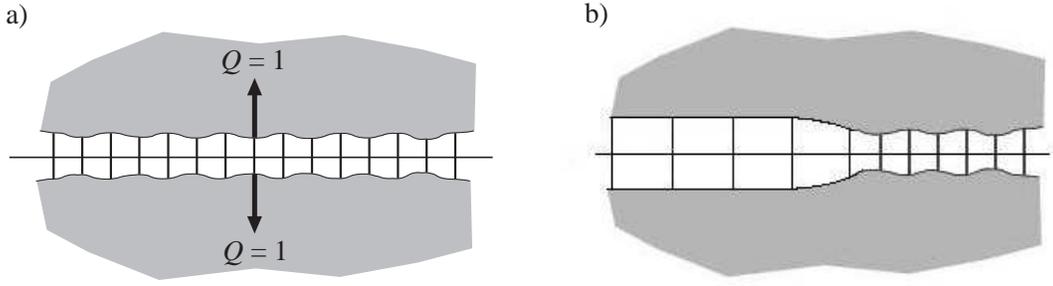}
    \caption{The body with the structured interface: (a) The Green's function related self-equilibrated couple of unit forces applied to the opposite ends of the bond at $m=0$. (b) The structure with a semi-infinite bridge crack, where the even bonds, $m=-2, -4, ...$, are broken.}
\label{f8}
\end{figure}

The difference in length results in a self-equilibrated stress
field that can cause a bridge crack with the even or odd bonds broken. As in the case of the two-line chain we assume without loss of generality that  the even bonds are broken. We also assume that,
since $|\GD|\ll h$,  the response of the structure to external forces corresponds to the regular, periodic set of the bonds. This also concerns the bulk of the body.

Specifically, we define the response by
the Green's function, $G(m)$, as the displacements at the upper end of the $m$-bond corresponding to a couple of self-equilibrated unit forces applied to the ends of the bond at the origin, $m=0$, \fig{f8}a.
For such periodic structure the displacement at the upper end of the $m$-bond  caused by a couple of forces $Q$ applied at $m=m'$ is
 \beq u_m = G(m-m')Q(m')\,.\eeq{1}
Note that if no external force acts on the half-planes except the Green's function source, the principle force acting on the upper half-plane from below is zero. It follows from the equilibrium condition that the applied unit force is completely supported by the bonds and hence
 \beq 2\Gk G^F(0) =1\,,\eeq{tfsbb}
where $(..)^F$ means the discrete Fourier transform
 \beq G^F(k)= \sum_{m=-\infty}^\infty G(m)\E^{\I k m}\,.\eeq{tfsbb1}

The half-planes are not stressed  if the forces equal to $\mp 2\Gk \GD$, are applied to the upper and lower ends of the odd bonds, respectively.  In this state, the position of the upper end of the bonds is the same as in the unstressed structure with the regular set of the bonds of the length $2h$: $y=h$. To remove these external forces we have to apply the opposite couple.

Thus, the displacements in the considered structure correspond to the regular structure under the couple of forces $\pm Q^0$ applied to the odd bonds, where
\beq Q^0  =2 \Gk \GD\,,\eeq{5}
and the positions of the upper end of the bonds are
 \beq y_m =h+ \sum_{m'=\pm 1, \pm 3, ...} G(m-m')Q^0(m')\,.\eeq{6}

Using the generalized Fourier transform we find the displacements from the $h$-level, $w_m=y_m-h$
 \beq w^F(k) = G^F(k)Q^0 \gl[\sum_{m=-\infty}^{-1}\exp((2m+1)(0+\I k))+\sum_{m=0}^\infty \exp(-(2m+1)(0-\I k))\gr]\n
 =  G^F(k)Q^0 \gl[\f{\exp(-(0+\I k))}{1-\exp(-(0+2\I k))} +\f{\exp(-(0-\I k))}{1-\exp(-(0-2\I k))}\gr]\n
 = \pi G^F(k)Q^0[\Gd(k)-\Gd(k-\pi)]\,, \eeq{7}
where $0\pm \I k$ is a limit of $\Ge \pm\I k$ as $0<\Ge \to 0$.

Using the inverse transform we find $w_m$ and the displacements from the initial sizes of the bonds, $\CU_m=w_m, m=0,\pm 2, \, ...$ , and $\CU_m=w_m-\GD, m=\pm 1, \pm3,\, ...$
 \beq  \CU_0=\f{1}{2} Q^0\gl[G^F(0)-G^F(\pi)\gr]~~~(m=0, \pm 2, ...)\,,\n
  \CU_1=  \f{1}{2} Q^0\gl[G^F(0)+G^F(\pi)\gr]-\GD~~~(m= \pm 1, \pm 3, ...)\,.\eeq{8}
Note that these relations correspond to the selective discrete transform (at $k=0$) introduced in the next section (also see Appendix).
The corresponding tensile forces in the bonds are
 \beq Q_{even}=  \Gk Q^0\gl[G^F(0)-G^F(\pi)\gr]~~~(m=0, \pm 2, ...)\,,\n
 Q_{odd}= \Gk\gl\{Q^0\gl[G^F(0)+G^F(\pi)\gr]-2\GD\gr\}\,.~~~(m= \pm 1, \pm 3, ...)\,.\eeq{9}
 With reference to Eqs. \eq{tfsbb}  and \eq{5}  we find that
 \beq Q_{odd}= - Q_{even}\,,~~~\CU_1=-\CU_0\eeq{10}
as it should be due to the equilibrium. Based on this and \eq{8}, \eq{10} another useful representation of $\CU_0$ can be obtained
 \beq \CU_0 = \GD\gl[1-\Gk\gl(G^F(0)+G^F(\pi)\gr)\gr]\,.\eeq{10a}
Note that $\GD$ is defined in \eq{9} if the tensile forces in the bonds are given.

\subsection{The static state with the semi-infinite bridged crack}
Let $\GD $ be positive, and the even bonds can be broken under extension. Consider the structure with the negative-number even bonds, $m=-2, -4, ...$, broken. The displacements additional to those in the initial state can be found considering the regular lattice loaded by external forces $Q^1= 2\Gk u_m, m=-2, -4, ...$
 (these external forces compensate the tensile forces in the corresponding bonds). The total displacements are
 \beq u_m = \CU_0 + U_m\,,~~~U_m = \sum_{m'= -2, -4,...}G(m-m') Q^1(m')~~~(m=0, \pm 2, ...)\,,\n
 u_m = \CU_1 + V_m\,,~~~V_m = \sum_{m'= -2, -4,...} G(m-m')Q^1(m')~~~(m=\pm 1, \pm 3, ...)\,,\eeq{ss1}
where the initial displacements, $\CU_0, \CU_1$, are defined in \eq{8}.

To proceed we introduce the following {\em selective} transforms (see also Appendix)
 \beq G_{even}^F(k) = \sum_{m=0, \pm 2, ...}G(m)\E^{\I k m}\,,~~~G_{odd}^F(k) = \sum_{m=\pm 1, \pm 3, ...}G(m)\E^{\I k m}\,,\n G_{even}^F(k)+G_{odd}^F(k)=G^F(k)\,.\eeq{ss2}
The selective transforms can be expressed through the regular one as follows
 \beq  G_{even}^F(k) = \f{1}{2}\gl[G^F(k)+G^F(k-\pi)\gr]\,,~~~G_{odd}^F(k) = \f{1}{2}\gl[G^F(k)-G^F(k-\pi)\gr]\,.\eeq{geo1}
Indeed, it can be seen that, in the inverse transform, the odd terms vanish for $ G_{even}^F(k)$ and  the even terms vanish for $ G_{odd}^F(k)$.

In terms of the corresponding transforms, the displacements additional to the initial state are
 \beq U^F(k)& =& 2\Gk G_{even}^F(k)\gl[U_-(k)+\f{\CU_0\E^{-2\I k}}{1-\E^{-2\I k-0}}\gr]\,,\n
 V^F(k)& =& 2\Gk G_{odd}^F(k)\gl[U_-(k)+\f{\CU_0\E^{-\I k}}{1-\E^{-2\I k-0}}\gr] \eeq{ss3}
with

 \beq U_-(k)=\sum_{m=-2, -4, ...}U_m\E^{\I k m}\,,~~~U_+(k)=\sum_{m=0, 2, ...}U_m\E^{\I k m}\,,\n U_-(k)+U_+(k)=U^F(k)\,. \eeq{ss4}

From the first relation in \eq{ss3} the Wiener-Hopf type equation follows as
 \beq U_+(k) + L(k)U_-(k) =\f{1-L(k)}{\E^{2\I k+0}-1}\CU_0\,,~~~
 L(k) = 1-2\Gk G_{even}^F(k)\,,\eeq{ss5}
and with reference to \eq{10a} and \eq{ss2}
 \beq \CU_0=\GD L(0)\,.\eeq{uoetdko}
Note that due to the symmetry $G_{even}^F(k)$ and hence $L(k)$ are even functions of real $k$. Besides, as is assumed here and below, the  kernel $L(k)>0$ on the real $k$-axis and can be factorised using Cauchy-type integral
 (in this connection see Slepyan, 2002, Sect. 12.2.3)
 \beq L(k)=\lim_{\Im k\to 0}L_+(k)L_-(k)\,,\n L_\pm(k) = \exp\gl[\pm\f{1}{4\pi\I}\int_{-\pi}^\pi \ln L(\xi)\cot \gl(\f{\xi-k}{2}\gr)\D \xi\gr]~~~(\pm\Im k >0)\,.\eeq{ss6}
In particular,
 \beq L_+(0)=L_-(0)=\sqrt{L(0)}\,,~~~L(\pm\pi)=L(0)\,,~~~L_+(\pm\pi)=L_-(\pm \pi)=\sqrt{L(0)}\,,\n
 L_+(\I\infty)=L_-(-\I\infty)=\exp\gl[\f{1}{2\pi}\int_{0}^\pi \ln L(\xi)\D \xi\gr]\,.\eeq{ss6p1}
Eq. \eq{ss5} can be rearranged as
 \beq \f{U_+(k)}{L_+(k)} + L_-(k)U_-(k) = \gl[\f{1}{L_+(k)}-L_-(k)\gr]\f{ \CU_0}{\E^{2\I k+0}-1}=C_++C_-\,,\n
 C_+= \gl[ \f{1}{L_+(k)}-\f{1}{L_+(0)}\gr]\f{\CU_0}{\E^{2\I k+0}-1}\,,~~~C_-=\gl[\f{1}{L_+(0)}-L_-(k)\gr]\f{\CU_0}{\E^{2\I k+0}-1}\,,\eeq{ss7}
where $C_+(C_-)$ has no singular and zero points in the upper (lower) half-plane of $k$.
We find
 \beq U_+(k) = C_+(k)L_+(k)\,,~~~U_-(k) = \f{C_-(k)}{L_-(k)}\,.\eeq{ss8}
The original functions can be obtained using the inverse transform formula
 \beq f_m=\f{1}{2\pi}\int_{-\pi}^\pi f^F(k)\E^{-\I k m}\D k\,,\eeq{ss8inv}
or a procedure of the consequent determination of the discrete terms (see Slepyan (2002), p.  65). In particular, the additional displacement at the crack front bond is
 \beq U_0 = \lim_{k\to \I\infty}U_+(k)=\f{\CU_0}{\sqrt{L(0)}}\exp\gl[\f{1}{2\pi}\int_{0}^\pi \ln L(\xi)\D \xi\gr]-\CU_0\eeq{ss9}
 with $\CU_0=\GD L(0)$.

\subsection{The energy relations}
We start from the state where the odd bonds are compressed by the force $Q^0=2\Gk\GD$. In this state, the strain energy per two spans is $2\Gk\GD^2$. Eliminating quasi-statically the compressive force we find the initial energy (per two span)
 \beq \CE = 2\Gk\GD^2 - Q^0(\GD-\CU_0)= 2\Gk\GD^2L(0)\,,\eeq{sse1}
 whereas the initial energy of a bond is $E=\CE L(0)$. The energy of the crack front bond is
 \beq E_0 = 2\Gk(\CU_0+U_0)^2=2\Gk\GD^2L(0)\exp\gl[\f{1}{\pi}\int_{0}^\pi \ln L(\xi)\D \xi\gr]\,.\eeq{sse2}
 The ratio $E_0/\CE$ is independent of the initial energy level. Thus, the energy release ratio is
 \beq R_\infty = \f{E_c}{\CE}= \exp\gl[\f{1}{\pi}\int_{0}^\pi \ln L(\xi)\D \xi\gr]\,.\eeq{mgf!}
It is remarkable that the ratio of the fracture energy to the totally released energy is given by this formula for a very general structure. The specific information is contained in the kernel of the Wiener-Hopf equation, $L(k)$, in the form given in \eq{ss5}. Note that this expression was first presented in Slepyan (1990) (also see Slepyan (2002)) in application to the lattice fracture under remote forces. Now it is found that the formula, in terms of $L(k)$, is quite general.

In the same way as for the chain in \az{lietlr}, it is found that for the general structure
 \beq \Gg=\Gg_c = \f{L(0)}{R_\infty}\,.\eeq{Gggen}

\subsection{Examples}
\subsubsection{The two line chain}\label{aie1}
We now return to the above-considered simple example. With the technique used in \az{aie} the corresponding Green's function can be found as
 \beq G(m)  = \f{\Gl^{|m|}}{2\Gm \sqrt{2\Ga+\Ga^2}}\eeq{ttlc1}
with
 \beq \Gl^2-2(1+\Ga)\Gl +1=0\,,~~~\Gl = 1+\Ga-\sqrt{2\Ga+\Ga^2}\,.\eeq{ttlc2}
Using the latter relations we find
 \beq G^F(k)=\f{1}{2\Gm(1+\Ga-\cos k)}\,,\n
 G_{even}^F(k) =\f{1+\Ga}{2\Gm[(1+\Ga)^2-\cos^2 k]}\,,~~~
 G_{odd}=\f{\cos k}{2\Gm[(1+\Ga)^2-\cos^2 k]}\,.\eeq{ss10}
It follows that
 \beq L(k) =1- 2\Gk G^F_{even}(k)=\f{1+\Ga-\cos^2 k}{(1+\Ga)^2-\cos^2 k} \,,\n
 L(0) = \f{1}{2+\Ga}\,,~~~ \CU_0= \f{\GD}{2+\Ga}\,.\eeq{ss11}

The latter expression coincides  with that in \eq{aie2a}. Calculations show that the other values, such as $U_0$ as a function of $\Ga$, also coincide with those obtained in \az{aie}.
It also can be found that in this case the expression for the initial energy in \eq{sse1} leads to that in \eq{ef6}.

 \subsubsection{Square-cell lattice}\label{scl}
\begin{figure}[h]
\begin{center}
\includegraphics[scale=0.50]{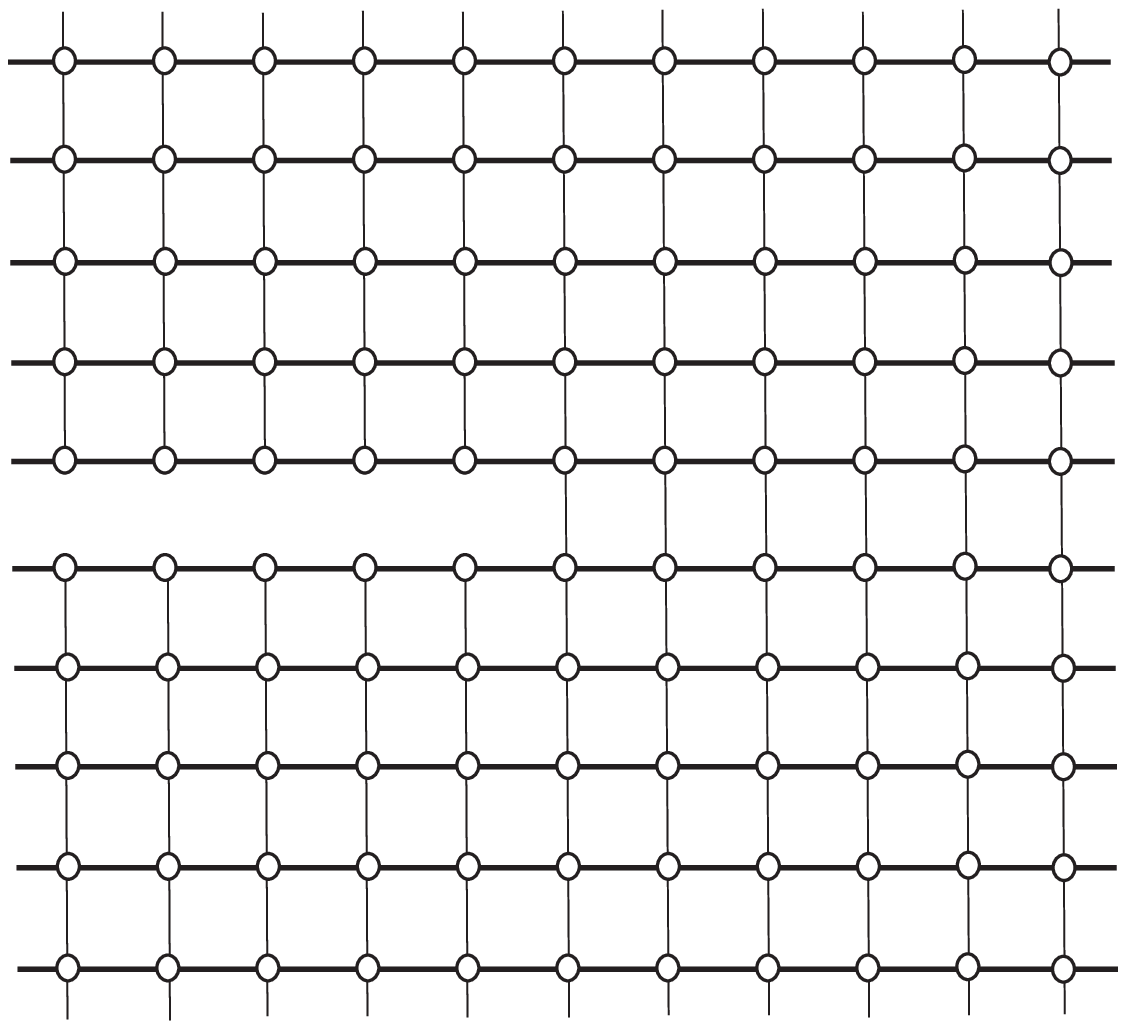}
\put(-173,-13){\small$m=-4$\hspace{0.5mm}$-3$\hspace{0.5mm}$-2$\hspace{0.5mm}$-1$%
    \hspace{2.7mm}$0$\hspace{3mm}$1$\hspace{3mm}$2$\hspace{3.2mm}$3$\hspace{3.3mm}$4$\hspace{3.3mm}$5$}
\put(-183,62){\small$-1$}\put(-195,78){\small$n=0$}\put(-175,93){\small$1$}\put(-175,108){\small$2$}\put(-175,123){\small$3$}
\put(-183,48){\small$-2$} \put(-183,34){\small$-3$}\put(-183,19){\small$-4$}
\end{center}
\caption{The orthotropic lattice. The stiffness of the horizontal and vertical bonds are $\Gm$ and $\Gk$, \res ($\Ga=\Gk/\Gm$). The bonds on the crack line are prestressed as in the chain.}
\label{f9}
\end{figure}

The equation for the square lattice, \fig{f9}, under the self-equilibrated couple of unit forces applied at $m=n=0$ (directed as the $y$-axis) and $m=0, n=-1$  (directed opposite) is
 \beq \Gm(2u_{m,n}-u_{m-1,n}-u_{m+1,n})+ \Gk(2u_{m,n}-u_{m,n-1}-u_{m,n+1})=\Gd_{m0}\Gd_{n0}-\Gd_{m0}\Gd_{n(-1)}\,,\eeq{scl1}
where $\Gm$ and $\Gk$  are stiffnesses of the horizontal and vertical bonds as for the two-line chain, and $u_{mn}$ is the displacement of the $m,n$ lattice node. Note that such anisotropic lattice has been considered in Mishuris et al, 2007; however, this was without evaluation of the Green's function.

Using the double Fourier discrete transform we find for the prospective crack face line, $n=0$
 \beq G^F(k,0)=u^F(k,0) = \f{1}{4\pi} \int_{-\pi}^\pi \f{1-\E^{-\I q}}{\Gm(1-\cos k)+\Gk(1-\cos q)}\D q \n =
 \f{1}{2\Gk}\gl(1- \f{1-\cos(k)}{\sqrt{(1-\cos(k))(1+2\Ga -\cos(k))}}\gr)\,.\eeq{sqcl1}
This is just what we need. Now, using relations in \eq{geo1} and \eq{ss5} we obtain
 \beq L(k)= \f{1}{2}\gl(\f{|\sin k/2|}{\sqrt{\Ga + \sin^2 k/2}}+\f{|\cos k/2|}{\sqrt{\Ga +\cos^2 k/2}}\gr)\,,~~~ L(0)=\f{1}{2\sqrt{\Ga+1}}\,,~~~ \CU_0=L(0)\GD\,.\eeq{sqcl2}

\begin{figure}[h]
\begin{center}
\includegraphics[scale=0.45]{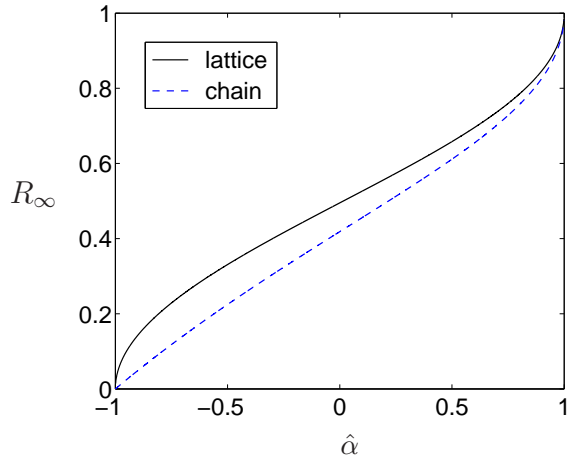}
  \put(-105,-5){\small$\hat\alpha$}
    \put(-230,90){$R_\infty$}
\end{center}
\caption{The energy ratio, $R_\infty$, as functions of  parameter
$\hat \alpha$ for the semi-infinite bridge crack in the orthotropic lattice and two line chain (the dotted line). }
\label{f10}
\end{figure}

It follows from \eq{Gggen} that limiting relations of $\Gg_c$ for the lattice with the semi-infinite bridge crack are
 \beq \lim_{\Ga\to 0} \Gg_c  = 1/2\,,~~~\lim_{\Ga\to\infty}\Gg_c \approx 0.78934782~~~(\lim_{\Ga\to\infty} R_\infty\approx 0.63343432)\,.\eeq{gamlimlat}
With reference to \eq{ss9}-\eq{sse2} and \eq{sqcl1}, \eq{sqcl2} the energy ratio $R_\infty $ follows as presented in \eq{mgf!} for a general case. It is plotted for the lattice and the chain in \fig{f10}. The plot of $\Gg_c(\hat{\Ga})$ is presented in \fig{Gg}.

\subsection{A finite crack}
In terms of the compensation forces, the displacements in the intact structure are as in \eq{1}, where
 \beq Q_m = 2\Gk\gl(\CU_{m} +U_m\gr)\eeq{afc1}
for $m\in M$, where $M$ is a set of numbers of the bonds which are expected to be broken. For the unknown displacements, $U_m$, we have the equations following from \eq{1}
 \beq U_m = 2\Gk\sum_{m\in M} G(m-m')\gl(\CU_{m'} +U(m')\gr)~~~(m\in M)\,.\eeq{afc2}

\section{The structure under combined actions of external forces and microstresses}
We consider two different loads: external forces, $q_0$, uniformly distributed at infinity and remote forces defined by their action at the crack tip with zero stresses at infinity. The uniform load is applicable in the case of a finite or semi-infinite bridged crack, whereas the latter type of the load corresponds to a semi-infinite crack without the crack-face bridging. Note that due to linearity the total displacements and tensile forces can be represented as  sums of those corresponding to the crack under the external load with no microstresses and to the crack under the microstresses without external loads. This superposition can be seen in the relations below.

\subsection{The bridged crack}
Consider the bridge crack shown in \fig{f8}b.
With respect to the determination of the tensile forces in the bonds, this state can be reflected by uniformly distributed self-equilibrated couples of forces, $q_0$, applied to the bonds. As a result, in the case of the semi-infinite bridged crack, the total displacement \eq{ss1} and the following relations including the final results \eq{ss8} and \eq{ss9} remain valid with the substitutions
  \beq u_{m} \longrightarrow u_{m}+q_0/(2\Gk)\,.\eeq{ss1ca1}

For the semi-infinite crack this formulation becomes invalid when $q_0$ reaches the value of a half of the critical tensile force in the odd number bonds. Indeed, at $m\to -\infty$ these bonds are loaded uniformly and they must carry the external load related to both the odd and the even bonds, while the latter are broken.

\subsection{The semi-infinite crack}
We now consider a general formulation for a semi-infinite crack under remote forces and microstresses. Let $m=-1$ be the number of the crack front bond. The bonds at $m=-2, -3, ...$ are assumed to be broken. As before we consider the intact structure under external forces, which compensate the tensile forces in the bonds at $m\le -$2. Thus, the additional displacements satisfy the equation
 \beq U_m = 2\Gk \sum_{m'=-\infty}^{-2} G(m-m') [ U(m') + (-1)^{m'}\CU_0]\,,\n
 U^F(k) = 2\Gk G^F(k) \gl[U_-(k) + \CU_0\sum_{m'=-\infty}^{-2}\E^{\I (k+\pi) m'}\gr]\n = 2\Gk G^F(k) \gl[U_-(k) + \CU_0\f{\E^{-2\I k}}{1
 +\E^{-\I k-0}}\gr]\,.\eeq{casic1}
Here
 \beq U_-(k)=\sum_{m'=-\infty}^{-2} U_m\E^{\I k m}\,,~~~U_+(k)=\sum_{m'=-1}^{\infty} U_m\E^{\I k m}\,.\eeq{casic1d}

 From \eq{casic1} we obtain the Wiener-Hopf type equation
 \beq U_+(k)+L(k)U_-(k)=2\Gk G^F(k) \CU_0\f{\E^{-2\I k}}{1
 +\E^{-\I k-0}}\,,~~~L(k)=1-2\Gk G^F(k)\,,~~~L(0)=0\,.\eeq{casic2}
We have to stress that $L(k)$ defined in this section in \eq{casic2} differs from that defined above in \eq{ss5}. Indeed, here we consider a fully opened semi-infinite crack (not the bridged crack as above).

After the factorization as in \eq{ss6} we represent this equation as
 \beq \f{U_+(k)}{L_+(k)} + L_-(k)U_-(k) = C\gl[\f{\E^{-\I k}}{1-\E^{\I k-0}} + \f{\E^{-2\I k}}{1-\E^{-\I k-0}}\gr]+S(k)\,,\eeq{suca2}
where the first term in the right-hand side reflects the remote forces, and
 \beq S(k)=\gl[\f{1}{L_+(k)}-L_-(k)\gr]\f{\CU_0\E^{-2\I k}}{1+\E^{-\I k-0}}
 =S_+(k)+S_-(k)\,,\n S_+(k)=\gl[\f{1}{L_+(k)}-\f{1}{L_+(\pi)}\gr]\f{\CU_0\E^{-2\I k}}{1+\E^{-\I k-0}}\,,~~~S_-(k)=\gl[\f{1}{L_+(\pi)}-L_-(k)\gr]\f{\CU_0\E^{-2\I k}}{1+\E^{-\I k-0}}\,.\eeq{suca2a}
The constant $C$ is determined in \az{caer} through the macrolevel energy release rate.

 It follows from these equations that
 \beq U_+(k)= L_+(k)\gl[\f{C\E^{-\I k}}{1-\E^{\I k-0}}+S_+(k)\gr]\,,~~~U_-(k)= \f{1}{L_-(k)}\gl[\f{C\E^{-2\I k}}{1-\E^{-\I k-0}}+S_-(k)\gr]\,,\eeq{suca3}
where, in particular, $L_+(k)$ and the first two terms in its representation as a Fourier series are ($\Im k >0$)
 \beq  L_+(k) = \exp\gl[\f{1}{4\pi\I}\int_{-\pi}^\pi \ln L(\xi)\cot \gl(\f{\xi-k}{2}\gr)\D \xi\gr] = \GF_1 +\GF_1\GF_2\E^{\I k} + ...\,,\n
L_+(\pi)=\sqrt{L(\pi)}\,,~~~ \GF_1=  \exp\gl[\f{1}{2\pi}\int_{0}^\pi \ln L(\xi)\D \xi\gr]\,,~~~\GF_2 =\f{1}{\pi}\int_{0}^\pi \ln L(\xi)\cos \xi \D \xi\,. \eeq{suca4}

Based on Eqs. \eq{suca2a} $-$ \eq{suca4} we determine explicit expressions for the first two coefficients in the series \eq{casic1d} for $U_+(k)$. Thereby we find the additional displacements caused by the crack
 \beq
 U(-1) = \GF_1 C +\gl(1-\f{\GF_1}{\sqrt{L(\pi)}}\gr)\CU_0\,,~~~\GF_1>0\,,\n
 U_0 = \GF_1(1+\GF_2)C - \gl(1-\f{\GF_1(1-\GF_2)}{\sqrt{L(\pi)}}\gr)\CU_0\,,~~~\GF_2<0\,.\eeq{suca5}
With account of the initial displacements the tensile forces in these bonds, the crack front bond initially compressed and the next one initially stretched, are
 \beq Q_{-1} = 2\Gk[U(-1)-\CU_0] = 2\Gk \GF_1\gl(C - \f{\CU_0}{\sqrt{L(\pi)}}\gr)\,,\n
 Q_0 = 2\Gk[U_0+\CU_0] = 2\Gk \GF_1\gl((1+\GF_2)C +\f{1-\GF_2}{\sqrt{L(\pi)}}\CU_0\gr)\,,\eeq{suca66}
where the initial displacement $\CU_0$ is defined in \eq{aie2a}.

\subsection{A finite crack}
For this problem, we use the equations in \eq{afc1} and \eq{afc2} adding to them the external forces
 \beq Q_m = 2\Gk\gl(\CU_m +U_m\gr)+ q_0\,.\eeq{afc11}
Thus, for the unknown displacements, $U_m$, we have the equations
 \beq U_m = \f{q_0}{2\Gk}+2\Gk \sum_{m'\in M} G(m-m')\gl(\CU_{m'} +U_m'\gr)~~~(m\in M)\,.\eeq{afc21}
The crack domain $M$ it can be found, in principle, using the strength criterion for the bonds and taking into account the fracture history. In this way, it can happen that there exists a bridge crack in a part of the total crack domain, where only even bonds are broken, and a free crack face sub-domain, where the odd bonds also are broken. Note, however, that dynamic effects can play an important role in the crack development.

\subsection{The energy release and the determination of the constant $\bfm{C}$}\label{caer}
Consider the long wave approximation of the Green's function and the kernel $L(k)$
 \beq G^F(k)\sim G_0^F(k)\,,~~~L(k)\sim L_0(k) ~~~(k\to 0)\,,\eeq{suca7}
 which correspond to the related continuum.
In this continuum, the stress on the crack continuation and the crack opening displacement have the following Fourier-transforms
 \beq \Gs_+(k) = \f{2\Gk L_{0+}(k) C}{a(0-\I k)}\,,~~~U_{-1}(k)= \f{C}{L_{0-}(k)(0+\I k)}\,,\eeq{suca8}
where $a$ is the distance between the neighboring bonds. In these terms, the energy release rate is (see Slepyan, 2002, p. 27 (1.42))
 \beq G_0= \lim_{s\to \infty}s^2 \Gs_+(\I s)U_{-1}(-\I s) = \f{2\Gk}{a} C^2 \lim_{s\to\infty}\f{L_{0+}(\I s)}{L_{0-}(-\I s)}\,.\eeq{suca9}
Since $L_{0\pm}(k)\to 1$ as $k\to \pm\I\infty$, \res, we obtain the connection between the macrolevel energy release rate and the constant $C$
 \beq G_{mac}= \f{2\Gk}{a} C^2\,.\eeq{suca10}
Note that $G_{mac}$ is the macrolevel energy release rate for the semi-infinite crack without bridging.

\subsection{Critical energy release rate and fracture scenarios}
With reference to \eq{sse1} and \eq{ss5} the microlevel energy release rate is
 \beq G_{mic}=\Gk\GD^2[1-2\Gk G^F(\pi)]/(2a)\eeq{mlerr}
and the total energy release rate is the sum
 \beq G_{tot}= G_{mac} + G_{mic}\,.\eeq{suca11}
In the case when the crack front bond brakes first, we have the following relation between the macro and micro parameters
 \beq E_{-1}=Q_{-1}[U(-1)-\CU_0]=2\Gk \GF_1^2\gl(C-\f{\CU_0}{\sqrt{L(\pi)}}\gr)^2=E_c\,.\eeq{suca12}
In the opposite case, when the next bond breaks first, the relation takes the form (see \eq{suca66})
 \beq E_0 =Q_0[U_0+\CU_0] = 2\Gk \GF_1^2\gl((1+\GF_2)C +\f{1-\GF_2}{\sqrt{L(\pi)}}\CU_0\gr)^2=E_c\,.\eeq{suca661}

Let the ratio of the initial energy of the bond to the critical energy be $\Gg$ as before
 \beq E = \Gg E_c~~(\CE=\Gg\CE_c)\,.\eeq{suca13}
With reference to \eq{suca10}, \eq{suca12} $-$ \eq{suca13}, taking the same critical force for both these bonds we find that the critical macrolevel energy release rate is
 \beq G_{mac}=\f{E_c}{a\GF_1^2}\gl(1+\GF_1\sqrt{\f{\Gg}{L(\pi)}}\gr)^2~~~(Q_{-1}=Q_c)\,,\n
 G_{mac}=\f{E_c}{a\GF_1^2(1+\GF_2)^2}\gl(1-\GF_1(1-\GF_2)\sqrt{\f{\Gg}{L(\pi)}}\gr)^2~~~(Q_{0}=Q_c)\,.\eeq{suca131}

\begin{figure}[h]
\begin{center}
\includegraphics[scale=0.45]{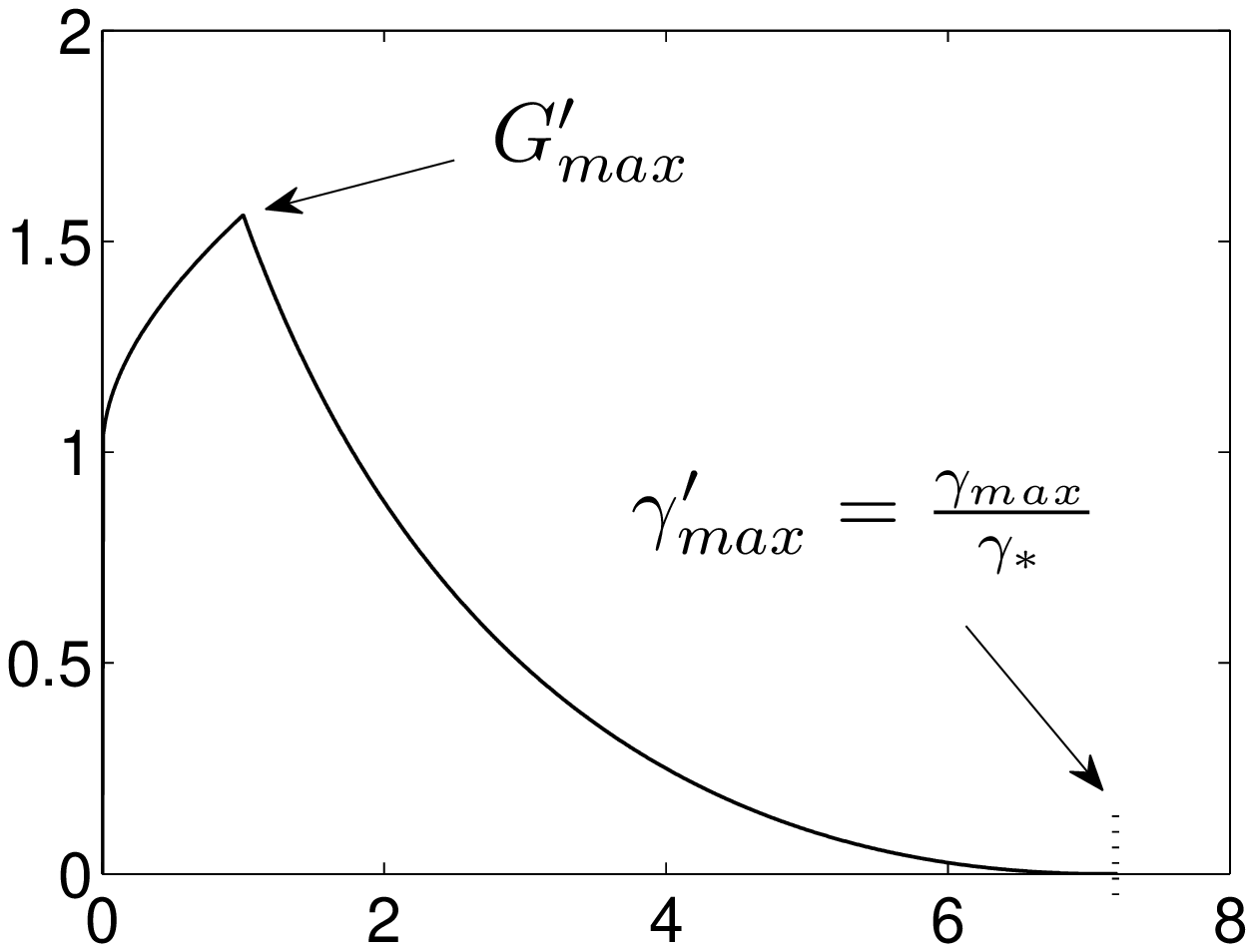}
  \put(-95,-10){\small$\gamma'$}
    \put(-230,80){$G'(\gamma')$}
\end{center}

\vspace*{-5mm}
\caption{The normalised macrolevel critical energy release rate, $G'$, as function of the normalised level of the internal energy, $\Gg'$. The equality $\Gg'=\Gg'_{max}$ corresponds to the critical level of the internal energy for the body with a semi-infinite open crack.}
\label{f11}
\end{figure}

Recall that $\GF_1>0, \GF_2<0$. The values defined by the above relations are equal at $\Gg=\Gg_*$, where
 \beq \Gg_*=L(\pi)\gl(\f{\GF_2}{2\GF_1}\gr)^2~~~\gl(\sqrt{\f{\Gg}{L(\pi)}}=- \f{\GF_2}{2\GF_1}>0\gr)\,.\eeq{suca77}

It follows that the crack front bond, $m=-1$, breaks first if $\Gg<\Gg_*$; otherwise, if $\Gg>\Gg_*$ the next bond, $m=0$, breaks first.
The normalized, critical, macrolevel energy release rate, $G'=G_{mac}a\GF_1^2/(E_c)$, as a function of $\Gg'=\Gg/\Gg_*$  (see \eq{suca131} and \eq{suca77}) becomes
  \beq G'=\gl(1-\f{\GF_2}{2}\sqrt{\Gg'}\gr)^2~~~(\Gg'<1)\,,\n
 G'=\f{1}{(1+\GF_2)^2}\gl(1+\f{\GF_2}{2}(1-\GF_2)\sqrt{\Gg'}\gr)^2~~~(\Gg'>1)\eeq{suca131a}
is presented in \fig{f11}. The plot is based on the relations in \eq{suca131} and \eq{suca77} with $\GF_2 = - 1/2$.
It can be seen that the crack-initiation threshold, which is the critical energy release rate required for the crack to advance, is an increasing function of the internal energy in the region, $0<\Gg<\Gg_*$. Then it decreases to zero in the remaining region, $\Gg_*<\Gg=1$.

Thus, different fracture scenarios are possible depending on the value of $\Gg$, that is, on the internal energy level. If $\Gg<\Gg_*$ a regular crack seems possible. Otherwise, a bridged crack region can occur, where only initially stretched bonds are broken. This region should increase with $\Gg$. A more detailed description of these scenarios is possible on the basis of the dynamic formulation.

\subsection{Concurrent fracture scenarios for a semi-infinite crack in the chain and the lattice under remote forces and microstresses}\label{CS}
With reference to \eq{ss10}, \eq{casic2} and \eq{suca4} it follows that for the chain
 \beq L(k) = \f{1-\cos k}{1+\Ga-\cos k}\,,~~~\GF_2=\Ga-\sqrt{2\Ga+\Ga^2}\,,\n
 \GF_1=\sqrt{1+\Ga/2}-\sqrt{\Ga/2}=\sqrt{1+\GF_2}\,,~~~\Gg_*=\f{\alpha}{2+\alpha}\,.\eeq{thtlcca1}

For the square lattice, as follows from \eq{sqcl1}  and \eq{casic2}
 \beq L(k) = \sqrt{\f{\sin^2 k/2}{\Ga+\sin^2 k/2}}\,, \quad \GF_2=-\sqrt{\Ga}(\sqrt{1+\Ga}-\sqrt{\Ga})\,, \n\GF_1=\sqrt{\sqrt{1+\Ga}-\sqrt{\Ga}}\,,\quad
 \Gg_*=\f{\alpha(\sqrt{1+\alpha}-\sqrt{\alpha})}{4\sqrt{1+\alpha}}\,.\eeq{LF1F2l}

\begin{figure}[h!]
    \hspace{10mm}\includegraphics [scale=0.50]{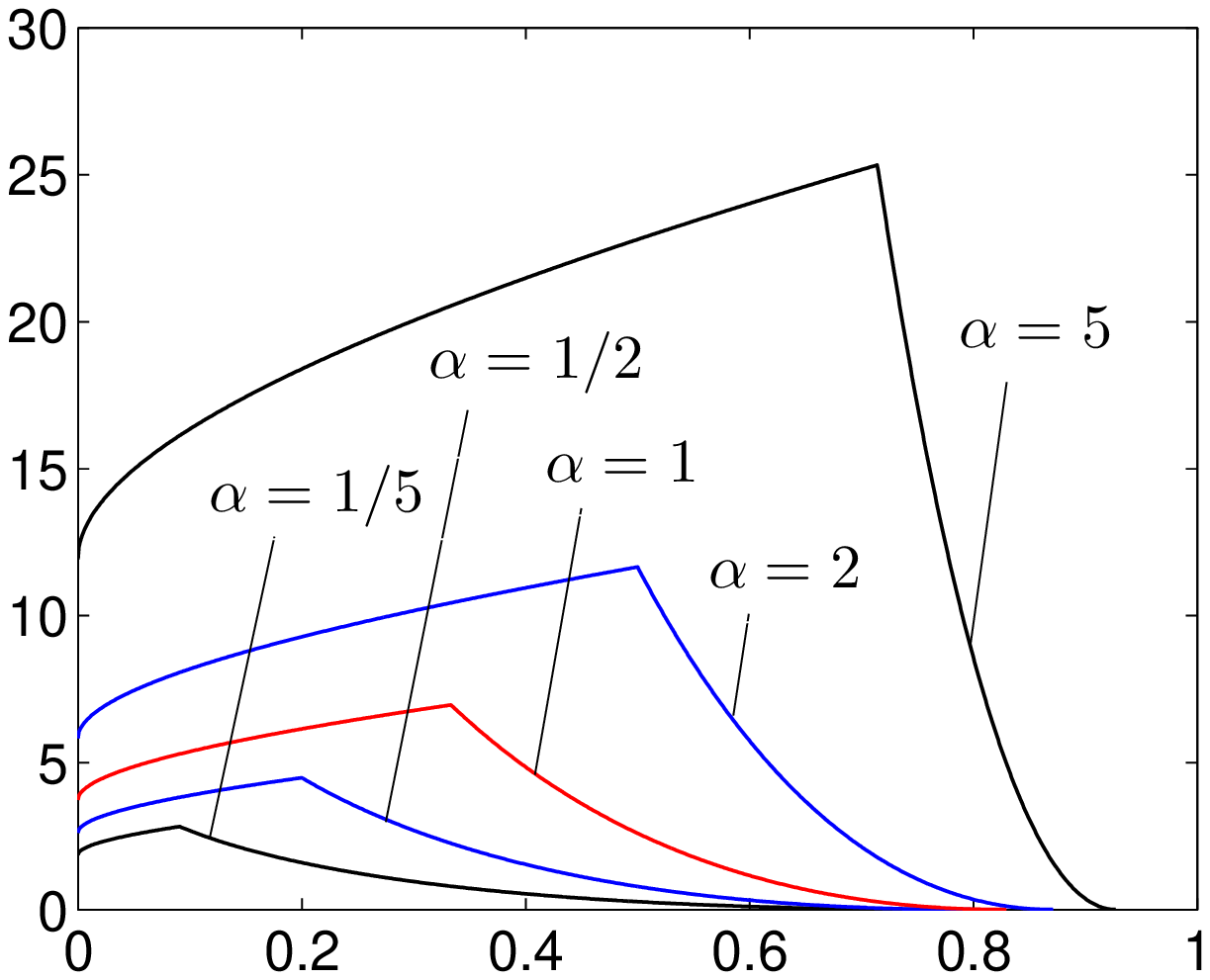}\hspace{5mm}
    \includegraphics [scale=0.50]{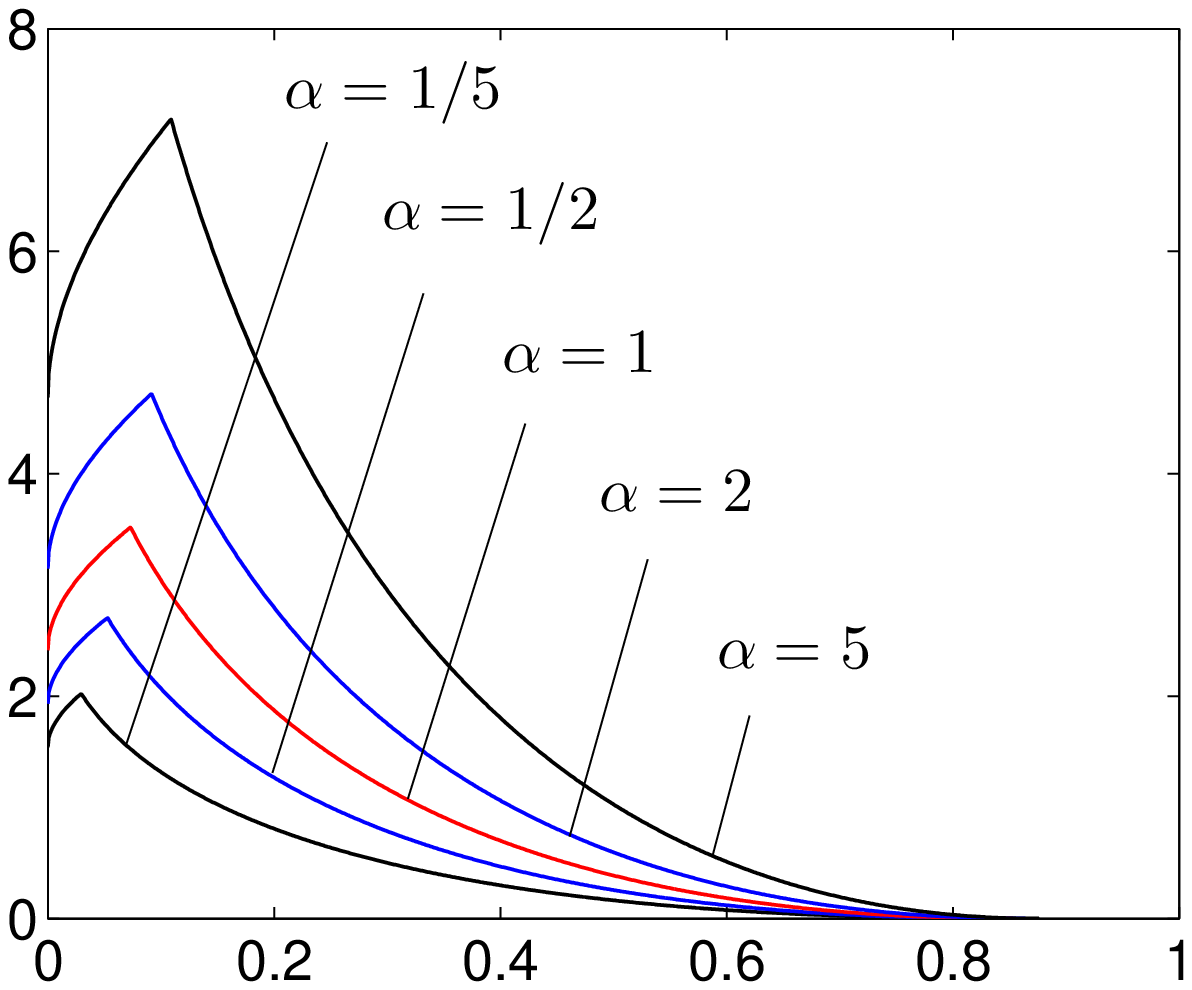}
\begin{picture}(-460,0)(0,0)
     \put(-340,-5){$\gamma$}
    \put(-110,-5){$\gamma$}
\put(-235,85){$\frac{aG_{mac}}{E_c}$}
\put(-460,85){$\frac{aG_{mac}}{E_c}$}
\put(-220,140){b)}
\put(-450,140){a)}
\end{picture}
\caption{The normalised, macrolevel, critical energy release rate, $aG_{mac}/E_c$, for the chain (a) and the lattice (b) as function of the normalised level of the internal energy, $\Gg$, for some values of $\Ga=\Gk/\Gm$.}
\label{f12}
\end{figure}

\begin{figure}[h!]
   \hspace{8mm} \includegraphics [scale=0.50]{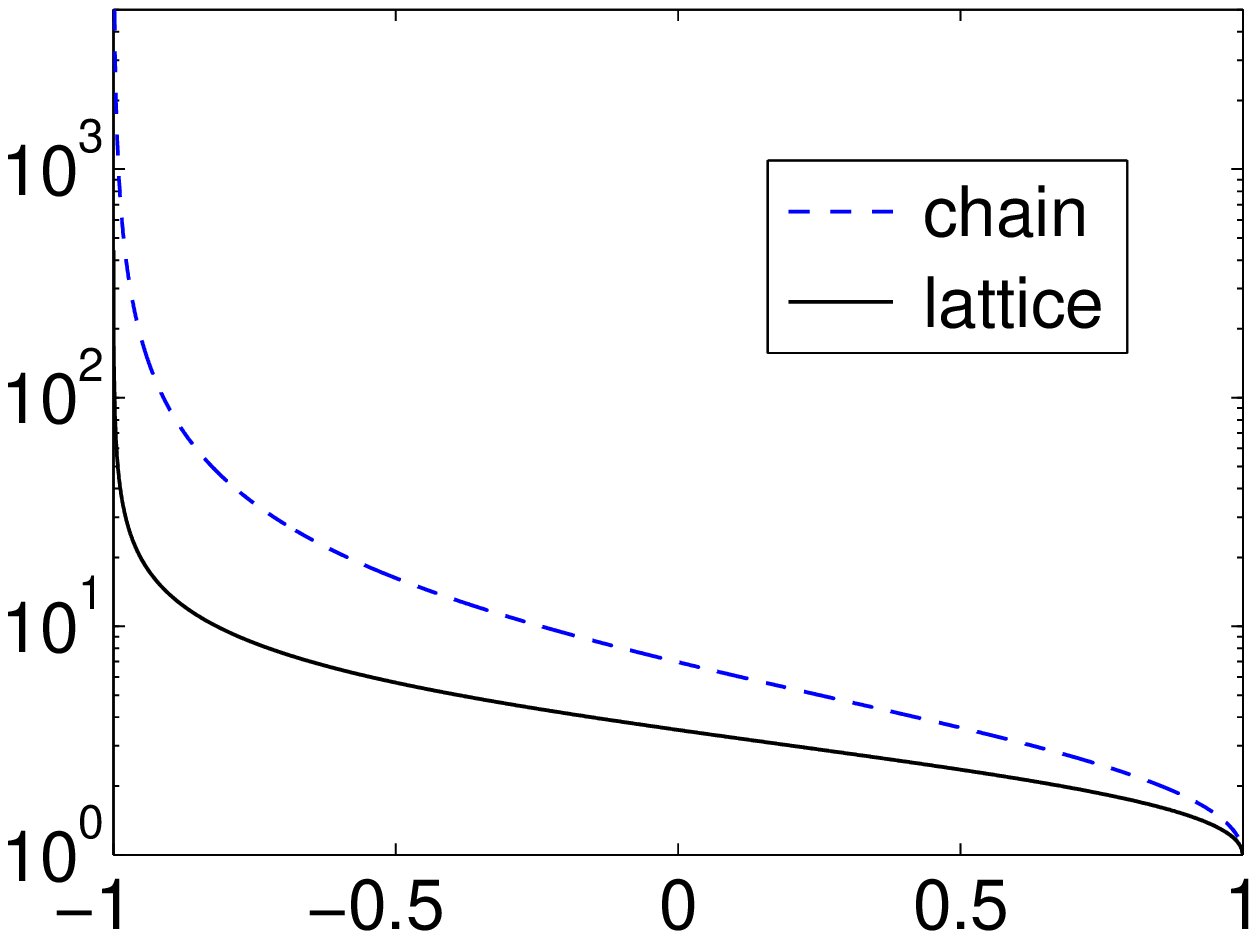}\hspace{8mm}\includegraphics [scale=0.50]{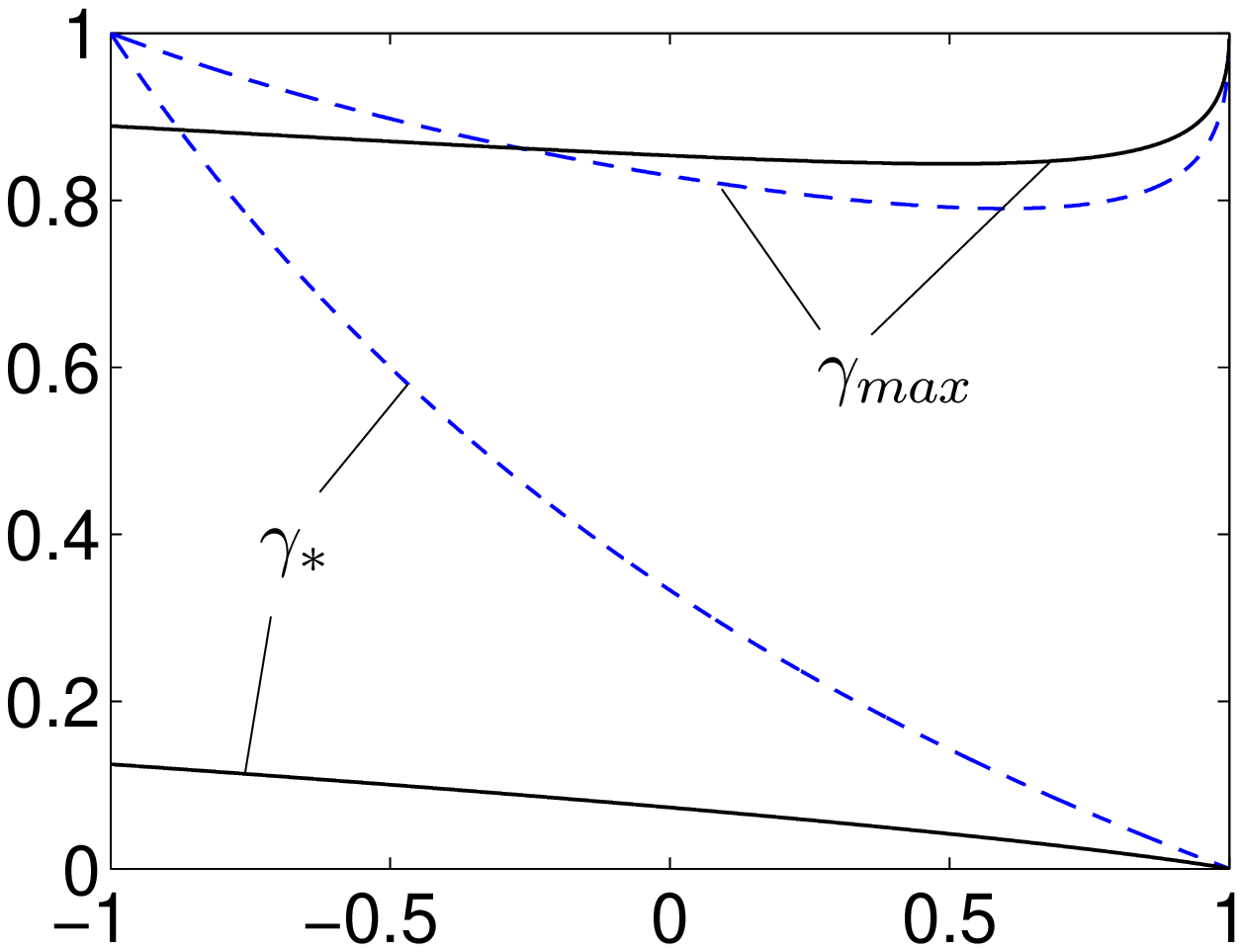}
\begin{picture}(-460,0)(0,0)
     \put(-345,-10){$\hat \alpha$}
    \put(-110,-10){$\hat \alpha$}

\put(-465,95){$\frac{aG_{mac}}{E_c}$}
\put(-220,140){b)}
\put(-430,140){a)}
\end{picture}
\vspace{3mm}

\caption{The normalized energy release ratio at $\Gg=\Gg_*$ (a) and the plots of $\Gg_*$  and $\Gg_{max}$ (b) as functions of $\hat{\Ga}=(1-\Ga)/(1+\Ga)$ for the chain and lattice. The equality $\Gg=\Gg_{max}$ corresponds to the critical level of the internal energy for the body with a semi-infinite open crack.}
\label{f13}
\end{figure}

The normalised, macrolevel, critical energy release rates, $aG_{mac}/E_c$, for the chain and lattice as functions of the normalised level of the internal energy, $\Gg$, is presented in \fig{f12} for some values of $\Ga=\Gk/\Gm$. Note that one point on these graphs, namely, the result for the isotropic lattice, $\Ga=1$, for the case of zero internal energy, $\Gg=0$, where $aG_{mac}/E_c=\sqrt{2}+1$, coincides with that found in Slepyan (1982). The values of $\Gg_*$ and $\Gg_{max}$ as functions of $\hat\Ga$ are plotted in \fig{f13}. Note that there is a large distance between these plots for the lattice. It follows that even for a low level of the internal energy, the second bond in front of the crack breaks first.

\vspace{5mm}

\vspace{5mm}

 \section{Discussions and Conclusions}
 In this paper, we have formulated a brittle fracture problem for a linearly elastic body with the internal potential energy of structure-associated alternating microstresses. The body of a non-specified structure is assumed to be periodic along an interface as the prospective crack path. In the general formulation, only the interface structure is specified. It is represented by a set of periodically placed bonds initially stressed due to their inconsistent strains. The initially stretched and compressed bonds alternate thus representing the self-equilibrated internal stress field. The stress-strain state of the interface, intact and with a crack, is examined, while the bulk of the body presence is reflected by a non-specified Green's function. In these terms, general solutions are obtained, and numerical results can be calculated as soon as the Green's function is specified. This is done with respect to a two-line chain and an anisotropic square lattice (the former is also considered independently). The crack equilibrium under microstresses and under a combined action of the internal and remote forces are examined.

 The results obtained evidence that, while the initial internal energy increases from zero to a threshold value, the crack initiation resistance also increases. With further increases in the internal energy, the resistance decreases to zero (see \fig{f11}, \fig{f12}). So, the microstresses can result in  an increase as well as in a decrease of the crack resistance depending on the internal energy level. The quasi-static considerations suggest that different scenarios for the crack growth can occur depending on the internal energy level. In particular, under a high level of microstresses the fracture can be accompanied by a bridged-crack region.

At least in the chain and lattice structures, a considerable part of the critical internal energy is radiated under the bond breakage (this part disappears in the framework of the quasi-static formulation). The remaining part of the released energy is the critical strain energy of the bond disappearing at fracture. The ratio of the breaking bond energy to the internal energy density for a finite crack rapidly approaches that for a semi-infinite crack, \fig{f6}, whereas the ratio to the total released energy is practically the same for any finite and semi-infinite cracks, \fig{f7}.

In analysis of the problem , a selective discrete transform is introduced (see \eq{ss2}, \eq{geo1} and Appendix). It is also found here that the formula for the ratio of the fracture energy to the total released energy \eq{mgf!}, obtained earlier for the lattice fracture under remote forces, is valid for a very general structure. The specific information is contained in the kernel of the Wiener-Hopf equation, $L(k)$, which has different expressions for the open and bridged semi-infinite cracks (under the same Green's function for the intact body).

In the paper, the interface structure is specified as a periodic set of the normally oriented bonds, \fig{f1}. The solutions are presented in terms of the displacements, which can also be interpreted as elongations. With this in mind we can conclude that all the considerations remain valid for such a set of regularly inclined bonds. The only condition is the uniformity of the Green's function with respect to the bond elongations. If this is the case, the bond elongations, caused by the interaction of the interface with the bulk of the body, can be represented by superposition based on the relation in \eq{1}.
In the case where, with respect to the structure elastic properties, the positive and negative $x$-directions are equivalent, a zigzag interface structure satisfies this condition. (In the latter, the bond inclination angles alternate but have the same absolute value as in the case of a triangular lattice.) Now the Green's function appearing in the obtained solutions must correspond to the couple of self-equilibrated unit forces oriented along the bond. Note that, in this case, not only fracture mode I or III can be considered as above but mode II as well.

As is common in the lattice fracture, the question remains how to homogenize the discrete model. In the framework of the present formulation, this question concerns only the interface. In a homogeneous material, the interface should have a zero thickness, whereas the alternating microstresses can be modelled by a sinusoidal self-equilibrated initial field. This can be imagined as a continuous elastic body comprised of two half-planes (or strips) with rough boundaries.  Physically such homogenization seems to be adequate; however, mathematically the problem becomes more complicated especially for the crack propagation. At the same time, the use of numerical methods returns us to one or the other discrete structure.

Finally, note that, under the quasi-static considerations adopted in this paper, we have determined the influence of the microstresses on the crack initiation criterion. Also the results allow us to predict some characteristics of the crack growth mode. However, the crack advance in a  perfect, brittle, discrete structure is accompanied by dynamic effects even for any small crack speed. Thus, to trace the crack growth, only the dynamic formulation is adequate. This is to be presented in separate publications, where we consider the spontaneous, like a domino wave, crack propagation under the internal energy and the crack dynamics under the combined action of the remote forces and microstresses. Along with the dynamic effects predicted in this paper a number of other dynamic phenomena revealed are discussed, such as hypersonic and quasi-stable slow cracks, clustering and variable finite bridging zones.

\vspace{3mm}
\noindent The authors {\bf acknowledge} support from the FP7 Marie Curie grant  No. 284544-PARM2.

\newpage

\vskip 18pt
\begin{center}
{\bf  References}
\end{center}
\vskip 3pt

\inh Banks-Sills, L., Ashkenazi, D., Eliasi, R., 1997. Determination of the effect of residual curing stresses
on an interface crack by means of the weight function method. Computational Mechanics 19, 507-510.

\inh Banks-Sills, L., Freed, Y., Eliasi, R., Fourman, V,. 2006. Fracture toughness of the $+45^o/ − 45^o$ interface of a laminate composite.
Int J Fract 141, 195-210.

\inh Barenblatt, G.I., 1959a.  The Formation of Equilibrium Cracks During Brittle Fracture: General Ideas and Hypotheses, Axially Symmetric Cracks. Appl Math Mech (PMM) 23, 622-636.

\inh Barenblatt, G.I., 1959b.  Concerning Equilibrium Cracks Forming During Brittle Fracture: The Stability of Isolated Cracks, Relationship with Energetic Theories. Appl Math Mech (PMM) 23, 1273-1282.

\inh Barenblatt, G. I., 1962. The mathematical theory of equilibrium cracks in brittle fracture. Advances in Applied Mechanics 7, 55-129.

\inh Bebamzadeh ,A., Haukaas, T.,  Vaziri, R.  Poursartip, A.,  Fernlund, G., 2009. Response Sensitivity and Parameter Importance in Composites Manufacturing. Journal of Composite Materials 43, 621-659.

\inh  Bebamzadeh, A., Haukaas, T.,  Vaziri, R., Poursartip, A., Fernlund, G., 2010. Application of Response Sensitivity in Composite Processing. Journal of Composite Materials 44, 1821-1840.

\inh Griffith, A.A., 1920. The Phenomena of Rupture and Flow in Solids. Phil Trans Roy Soc (London) A221, 162-198.

\inh Griffith, A.A., 1924. The theory of rupture. Proc. First Inter. Congress Appl. Mech., Delft, 55-63.

\inh Mishuris, G.S., Movchan, A.B., Slepyan, L.I., 2007. Waves and fracture in an inhomogeneous lattice structure. Waves in Random and Complex Media, 17, 409-428.

\inh  Mishuris, G.S., Movchan, A.B., Slepyan, L.I.,  2008. Dynamics of a bridged crack in a discrete lattice. The Quarterly Journal of
Mechanics and Applied Mathematics, 61, 151-160.

\inh Mishuris, G.S., Movchan, A.B., Slepyan, L.I., 2009. Localised knife waves in a structured interface. J. Mech. Phys. Solids 57, 1958-1979.

\inh Mishuris, G.S., Movchan, A.B. and Bigoni, D. 2012. Dynamics of a fault steadily propagating within a structural interface.
SIAM Journal on Multiscale Modelling and Simulation, 10(3), 936-953.

\inh Novozhilov, V.V., 1969. On a necessary and sufficient criterion for brittle strength. PMM  33(2),  212-222.

\inh Novozhilov, V.V., 1969. On the foundations of a theory of equilibrium cracks in elastic solids. PM M 33(5), 797-812.

\inh  Qi Zhu,   Philippe H. Geubelle,   Min Li,    Charles L. Tucker III, 2001. Dimensional Accuracy of Thermoset Composites: Simulation of Process-Induced Residual Stresses. Journal of Composite Materials 35, 2171-2205.

\inh Slepyan, L.I., 1981a. Dynamics of a crack in a lattice. Sov. Phys. Dokl., 26, 538-540.

\inh Slepyan, L.I., 1981b. Mechanics of cracks. Sudostroenie, Leningrad.

\inh Slepyan, L.I., 1982. Antiplane Problem of a Crack in a Lattice. Mechanics of Solids, 17, 101-114.

\inh Slepyan, L.I., 1990. Mechanics of cracks. Sudostroenie, Leningrad (2-nd edition).

\inh Slepyan, L.I., 2000. Dynamic Factor in Impact, Phase Transition and Fracture. J. Mech. Phys. Solids, 48, 927-960.

\inh Slepyan, L.I., and Ayzenberg-Stepanenko, M.V., 2002. Some surprising phenomena in weak-bond fracture of a triangular lattice.
J. Mech. Phys. Solids 50(8), 1591-1625.

\inh Slepyan, L.I., 2002. Models and Phenomena in Fracture Mechanics. Springer, Berlin.

\inh Slepyan, L.I., 2010a. On discrete models in fracture mechanics. Mechanics of Solids 45(6), 803-814.

\inh Slepyan, L.I., 2010b. Wave radiation in lattice fracture. Acoustical Physics, 56(6), 962-971.

\inh Slepyan, L.I., Mishuris, G.S., Movchan, A.B., 2010.  Crack in a lattice waveguide. Int J Fract 162, 91-106.

\inh Thompson, Robb, Hsieh, C. and Rana, V., 1971. Lattice trapping of fracture cracks. J. Appl. Phys.,  42, 3154-3160.

\inh White, S.R., and Hahn, H.T., 1992. Process Modeling of Composite Materials: Residual Stress Development during Cure. Part I. Model Formulation.  Journal of Composite Materials January  26, 2402-2422.

\inh White, S.R., and Hahn, H.T., 1992. Process Modeling of Composite Materials: Residual Stress Development during Cure. Part II. Experimental Validation. Journal of Composite Materials  26, 2423-2453.

\inh Willis, J. R., 1967. A comparison of the fracture criteria of Griffith and Barenblatt.  J. Mech. Phys. Solids 15(3), 151-162.

\section*{Appendix: Selective discrete transform}

\appendix

Along with the regular discrete Fourier transform
 \beq f^F(k)= \sum_{m=-\infty}^\infty f(m)\E^{\I km}\eeq{sdt1}
consider a \ind{selective discrete transform}
 \beq f_{n\nu}^F(k) = \sum_{\Gm=-\infty}^\infty f(\nu +\Gm n)\E^{\I k(\nu+\Gm n)}\,,\eeq{sdt2}
 where $\nu$ and $n$ are integers, $0\le \nu\le n-1, n\ge 1$.

The statement is true that the latter can be expressed in terms of the former as follows
 \beq    f_{n\nu}^F(k) =\f{1}{n} \sum_{m=0}^{n-1}\exp\gl(\f{2\pi\I m\nu}{n}\gr) f^F\gl(k-\f{2\pi m}{n}\gr)\,.\eeq{sdt3}
Indeed, for any integer $\nu'$, $0\le \nu' \le n-1$, we find that the original function corresponding to this transform is
 \beq f_{n\nu}(\nu'+\Gm n)=\f{1}{n} \sum_{m=0}^{n-1}\exp\gl(\f{2\pi\I m\nu}{n}\gr)\f{1}{2\pi} \int_{-\pi}^\pi f^F\gl(k-\f{2\pi m}{n}\gr)\E^{-\I k(\nu'+\Gm n)}\D k\n
 = \f{1}{n} \sum_{m=0}^{n-1}\exp\gl[\f{2\pi\I m}{n}(\nu - \nu')\gr] \n
\times \f{1}{2\pi}\int_{-\pi}^\pi f^F\gl(k-\f{2\pi m}{n}\gr)\exp\gl[-\I \gl(k-\f{2\pi m}{n}\gr)(\nu'+\Gm n)\gr]\D k\n
 = f(\nu'+\Gm n)\f{1}{n}
 \sum_{m=0}^{n-1}\exp\gl[\f{2\pi\I m}{n}(\nu - \nu')\gr]\,.\eeq{sdt5}
Thus
 \beq f_{n\nu}(\nu'+\Gm n)= f(\nu+\Gm n)~~~(\nu'=\nu)\,,\n
  f_{n\nu}(\nu'+\Gm n)=  f(\nu+\Gm n) \f{1}{n}\f{1-\exp[2\pi\I(\nu-\nu')]}{1-\exp[2\pi\I(\nu-\nu')/n]}=0~~(\nu' \ne \nu)\,. \eeq{sdt6}
Note that we have used here the fact that the integrand is a $2\pi$-periodic function of $k$. Also note that a sum of the selective transform is the original transform
 \beq \sum_{\nu=0}^{n-1}f_{n\nu}^F(k)=f^F(k)\,.\eeq{stds1}

\end{document}